\documentclass[conference]{IEEEtran}
\pdfoutput=1

\usepackage{cite}
\usepackage{graphicx}
\usepackage{subcaption}
\usepackage{refstyle}
\usepackage[cmex10]{amsmath}
\usepackage{algorithmic}
\usepackage{array}
\usepackage{url}
\usepackage{fontenc}
\usepackage[utf8]{inputenc}
\usepackage{mathtools}
\usepackage{tabularx}
\usepackage{tabu}
\usepackage{float}
\usepackage{multirow}
\usepackage{siunitx}
\usepackage{algorithmic}
\usepackage{psfrag}
\usepackage{pgfplots}
\usepackage{xcolor}
\newcommand\note[1]{\textcolor{red}{#1}}
\usepackage{comment}
\usepackage{wrapfig}

\usepackage[utf8]{inputenc}
\usepackage{amsmath,amssymb}
\usepackage{enumitem}
\newcommand{\abbrev}{\texttt{Rand-OFDM}}

\newsavebox{\ieeealgbox}

\newcolumntype{M}[1]{>{\arraybackslash}m{#1}}
\newcolumntype{N}{@{}m{0pt}@{}}

\newtheorem{lemma}{Lemma}
\newtheorem{proof}{Proof}
\usepackage{proof}

\begin{document}


\title{{\abbrev}: A Secured Wireless Signal}

\author{\IEEEauthorblockN{Hesham Mohammed and Dola Saha}
\IEEEauthorblockA{\textit{Department of Electrical \& Computer Engineering} \\
\textit{University at Albany, SUNY}\\
\{hhussien, dsaha\} @albany.edu}
}

\IEEEoverridecommandlockouts
\maketitle
\pagestyle{plain}


\begin{abstract}

Wireless communication has been a broadcast system since its inception, which violates security and privacy issues at the physical layer between the intended transmit and receive pairs. As we move towards advanced spectrum sharing methodologies involving billions of devices connected over wireless networks, it is essential to secure the wireless signal such that only the intended receiver can realize the properties of the signal. In this paper, we propose {\abbrev}, a new waveform, secured with a shared secret key, where the signal properties can be recovered only at the expected receiver.
We achieve the signal level security by modifying the OFDM signal in time-domain, thus erasing the OFDM properties and in turn obfuscating the signal properties to an eavesdropper. We introduce a key-based secured training signal for channel estimation, which can be used only at the intended receiver with prior knowledge of the key. As a final step for the recovery of the signal, we use a clustering based technique to correct the phase of the received signal.
Our cryptanalysis shows that {\abbrev} is especially useful in future wide band signals. Extensive simulation and over-the-air experiments show that the performance of {\abbrev} is comparable to legacy OFDM and SNR penalty due to the secured waveform varies between $\approx$ 1-4dB. In all these scenarios, {\abbrev} remains unrecognized by the adversary even at the highest possible SNR.

\end{abstract}

\section{Introduction}

As new spectrum (sub-6GHz, mmWave and TeraHertz) becomes available for communication and coexistence of frequency-agile cognitive heterogeneous nodes becomes a norm, we need to rethink physical layer security to provide maximum secrecy in a broadcast channel. There has been a growing interest recently among multiple federal agencies to utilize the spectrum in a collegial way by heterogeneous devices. This also indicates that wireless signals will be vulnerable to various security attacks, which was unforeseen in prior extremely regulated framework. This motivates us to investigate in physical layer secured communication, which is hard to decipher by an eavesdropper in a hostile scenario, as well as practical enough to be accepted for mass deployments.

Prior work~\cite{Poor19, Survey, sdr_sec, sec_cogradio_13, sec_cogradio_15}
utilizes imperfection of the communication channel to establish secrecy by physical layer methods without the need of a shared secret key. However, channel imperfections are not enough to provide high secrecy capacity when the eavesdropper has a high signal-to-noise ratio (SNR) or has similar quantized channel state as the intended receiver. Higher layer encryption~\cite{sec_survey_16} provides computationally hard secrecy, but there exists key management and distribution issues. Also, since the waveform remains unchanged in all the above mentioned scenarios, it is plausible for the eavesdropper to restrict cryptanalysis search space within that waveform. To address these issues for future wireless agile radios, we introduce physical layer security, where we modify the OFDM waveform to completely disrupt its orthogonality properties of the subcarriers based on a shared secret key. At the receiver, we perform extensive channel estimation and reconstruct the waveform. The secret key can be derived from the channel or stored in the radio hardware, which will minimize the key distribution issues. In this paper, we only focus on modification of the time-domain signal at the transmitter and reconstruction of that back at the receiver.

\begin{figure}
    \centering
    \includegraphics[width=0.8\linewidth]{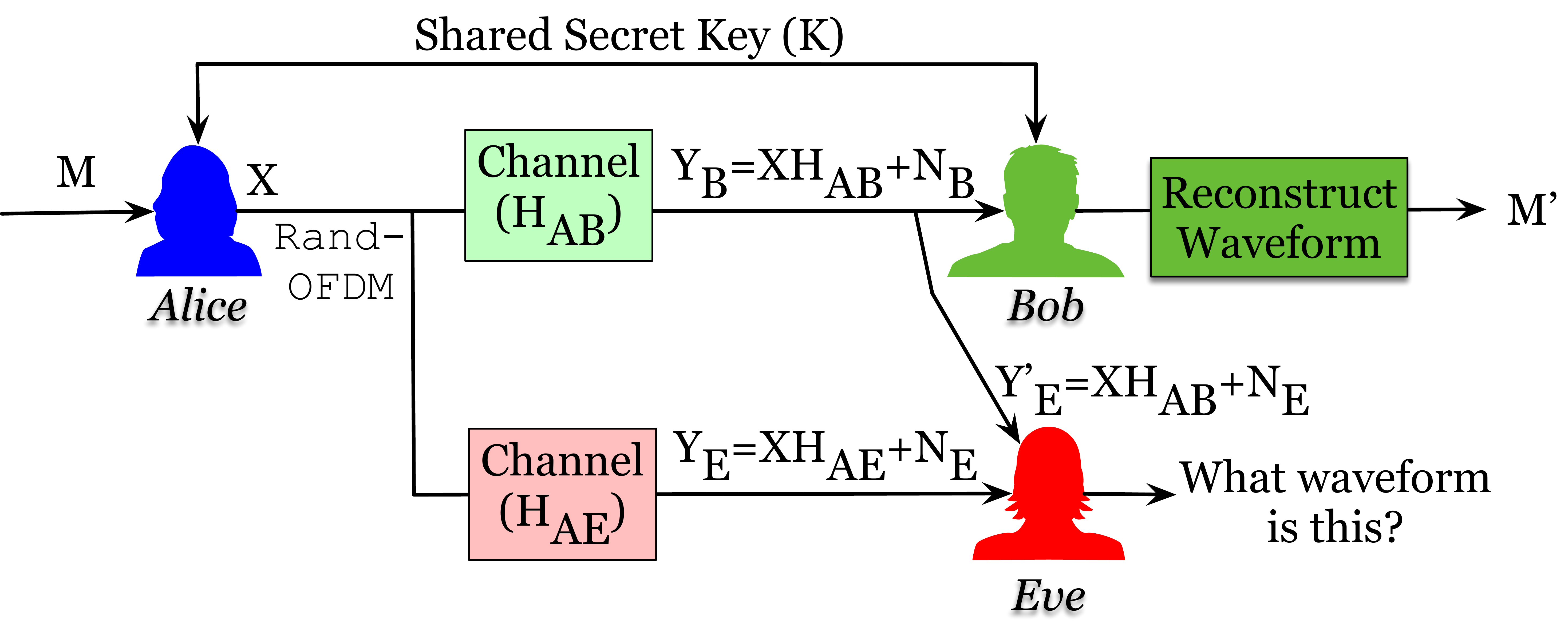}
    \vspace{-10pt}
    \caption{{\abbrev}: A system overview.}
    \label{fig:case}
\end{figure}

If $M$ is the message that \textit{Alice} needs to transmit to \textit{Bob}, she can encrypt it with the shared secret key ($K$) to produce $X$. As it reaches \textit{Bob}, it passes through the channel $H_{AB}$, such that the received signal is $Y_B=XH_{AB}+N_B$, where $N_B$ is the noise. An eavesdropper, \textit{Eve}, will experience a different channel and will receive $Y_E=XH_{AE}+N_E$. If channel imperfections are used to encrypt, secrecy capacity is a function of received SNR at \textit{Eve}. In this scenario, we propose {\abbrev}, where we modify the time domain signal, such that the transmitted signal, $X$, is no longer a known waveform (OFDM, CDMA, etc.). Figure~\ref{fig:case} shows an overview of {\abbrev}, where we randomize the time domain OFDM signal based on $K$, such that the resultant $X$ does not have orthogonality property. Due to the spectral efficiency of OFDM and its use in most standards, like Wi-Fi~\cite{802_11_spec}, 5G~\cite{lte_book}, we have chosen it as the candidate for generating the base waveform.

Our approach has two distinct benefits over prior work. {\bf \textit{Firstly}}, secret key based approach to modify the signal indicates that even at high SNR or even if \textit{Eve} has full channel knowledge between \textit{Alice} and \textit{Bob}, $H_{AB}$, she will not be able to decode $X$. In other words, if \textit{Eve} gets access to $Y_{E}'=XH_{AB}+N_B$, she will not be able to decode $X$ without the key $K$, which is used to generate $X$. {\bf \textit{Secondly}}, by modifying the time domain signal based on a key, we ensure that it appears as a noise or an unknown signal to \textit{Eve}. The combination of the key and data will create a different signal every time it is transmitted. Thus it creates a larger search space for \textit{Eve} to attack this waveform. It is to be noted here that any higher layer encryption or bit level interleaving can be used in conjunction with {\abbrev} to provide multiple layers of protection from different security threats.

The major challenges of {\abbrev} are: 1) to design a computationally light operation to modify the OFDM waveform in time-domain, 2) to estimate the channel effects accurately at the receiver, which looses frequency domain properties due to the time domain modifications, 3) to hide the channel parameters in a way to make it difficult for the eavesdropper to estimate the channel accurately, and 4) to architect a design to correct the residual phase offsets due to the absence of pilot subcarriers in frequency domain. 
To the best of our knowledge, we are the first ones to \textit{modify the time domain wireless OFDM signal to loose the orthogonality of the signal and introduce novel channel estimation technique to recover the signal back at the receiver and evaluated the system in practical over-the-air scenarios}. Hence, the key contributions of our work can be listed as follows:

 \begin{figure*}[h]
    \centering
    \begin{subfigure}[b]{0.24\textwidth}
        \includegraphics[width=\textwidth]{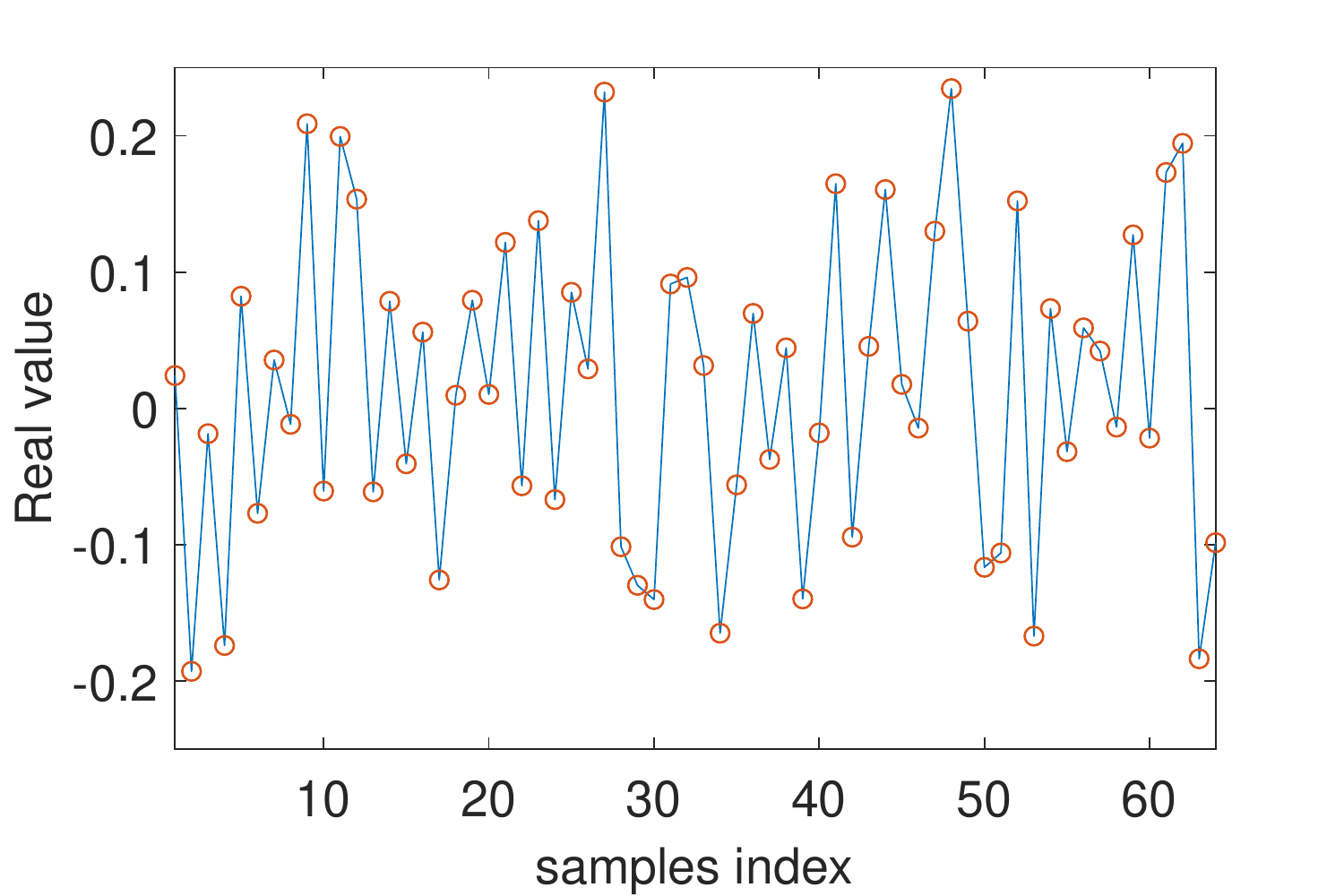}
        \caption{Time domain OFDM signal before randomization}
        \label{fig:ofdm}
    \end{subfigure}
    \begin{subfigure}[b]{0.24\textwidth}
        \includegraphics[width=\textwidth]{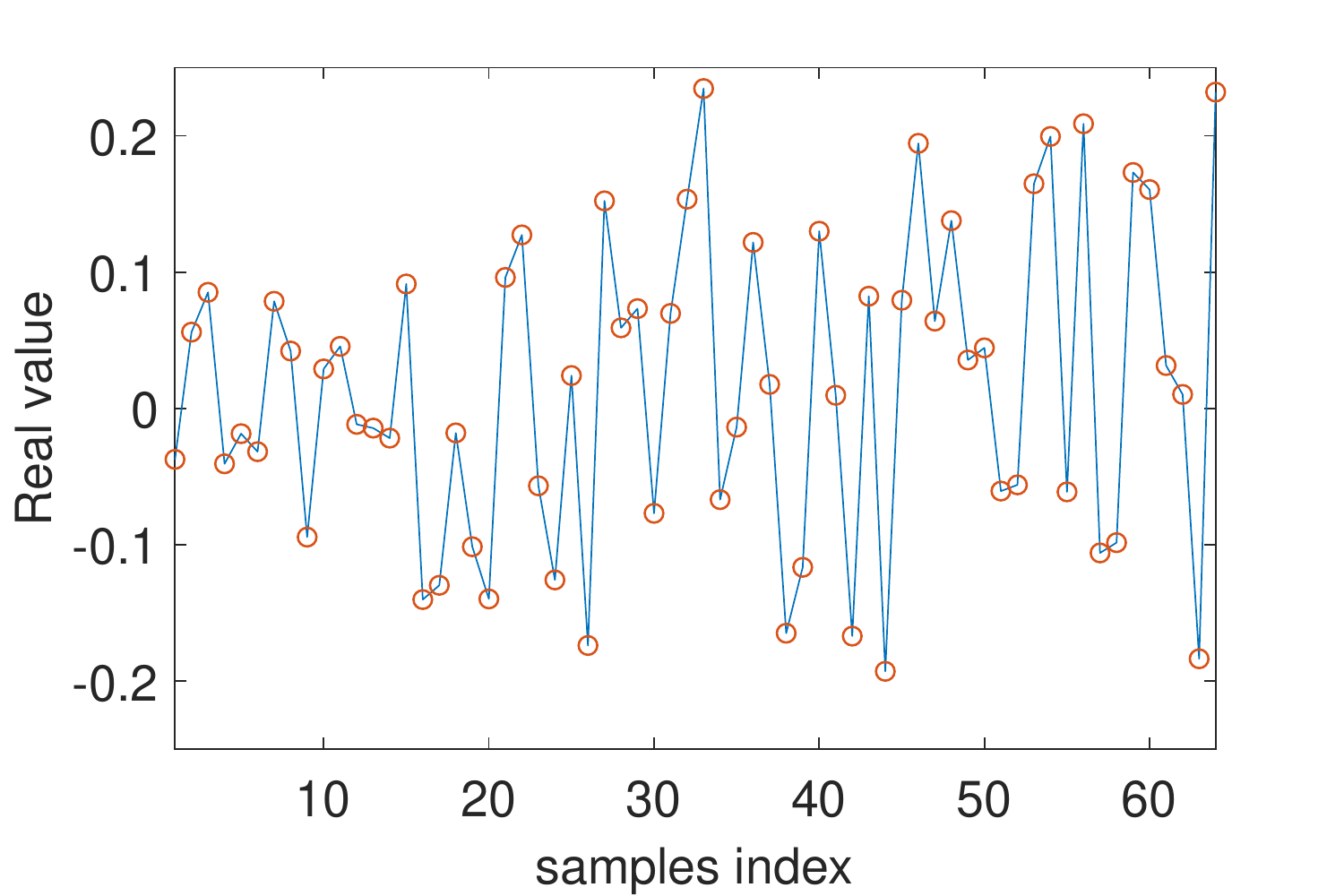}
        \caption{Time domain {\abbrev} signal after randomization}
        \label{fig:rand_ofdm}
    \end{subfigure}
    \begin{subfigure}[b]{0.24\textwidth}
        \includegraphics[width=\textwidth]{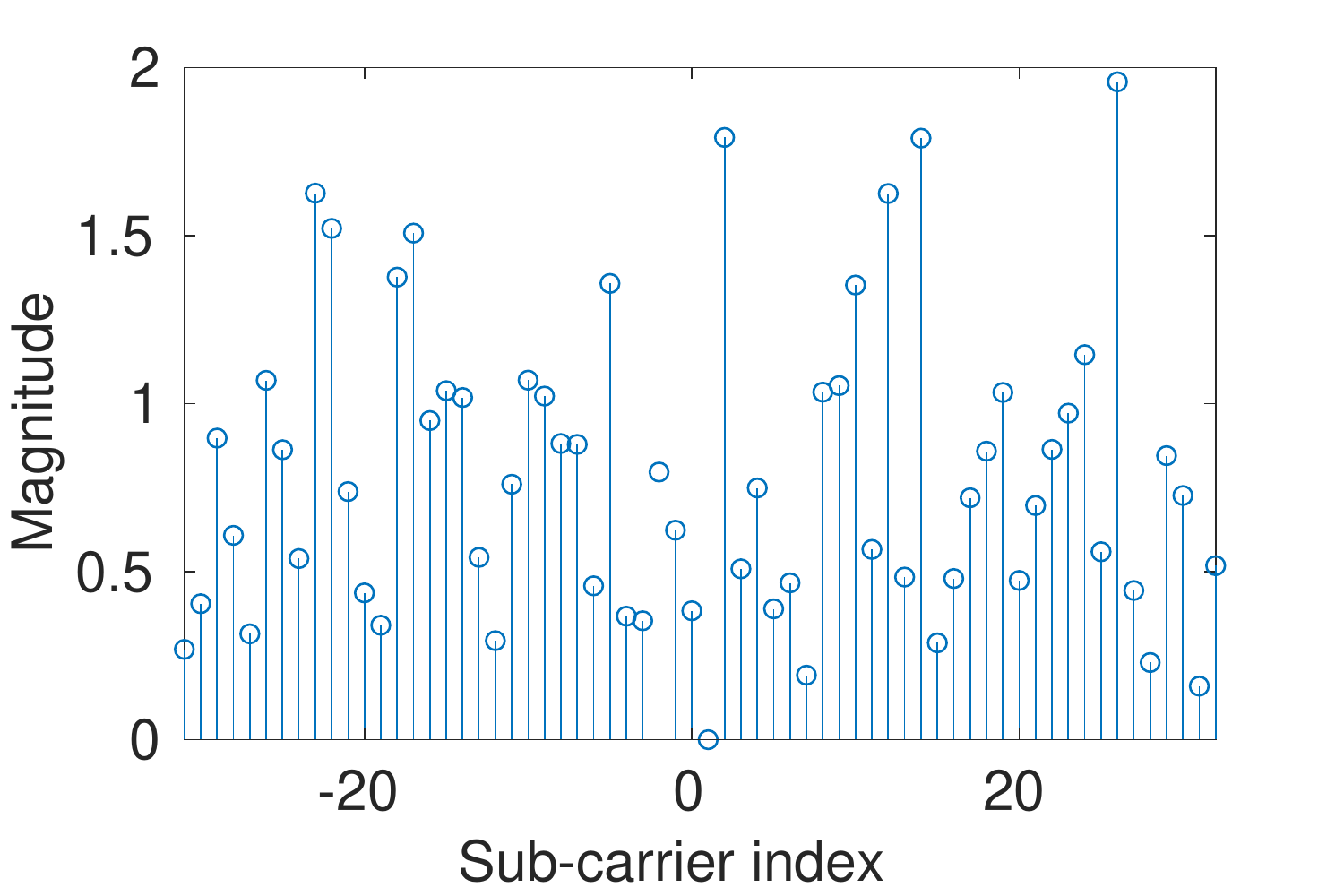}
        \caption{Frequency response of {\abbrev} signal}
        \label{fig:rand_ofdm_fft}
    \end{subfigure}
    \begin{subfigure}[b]{0.24\textwidth}
        \includegraphics[width=\textwidth]{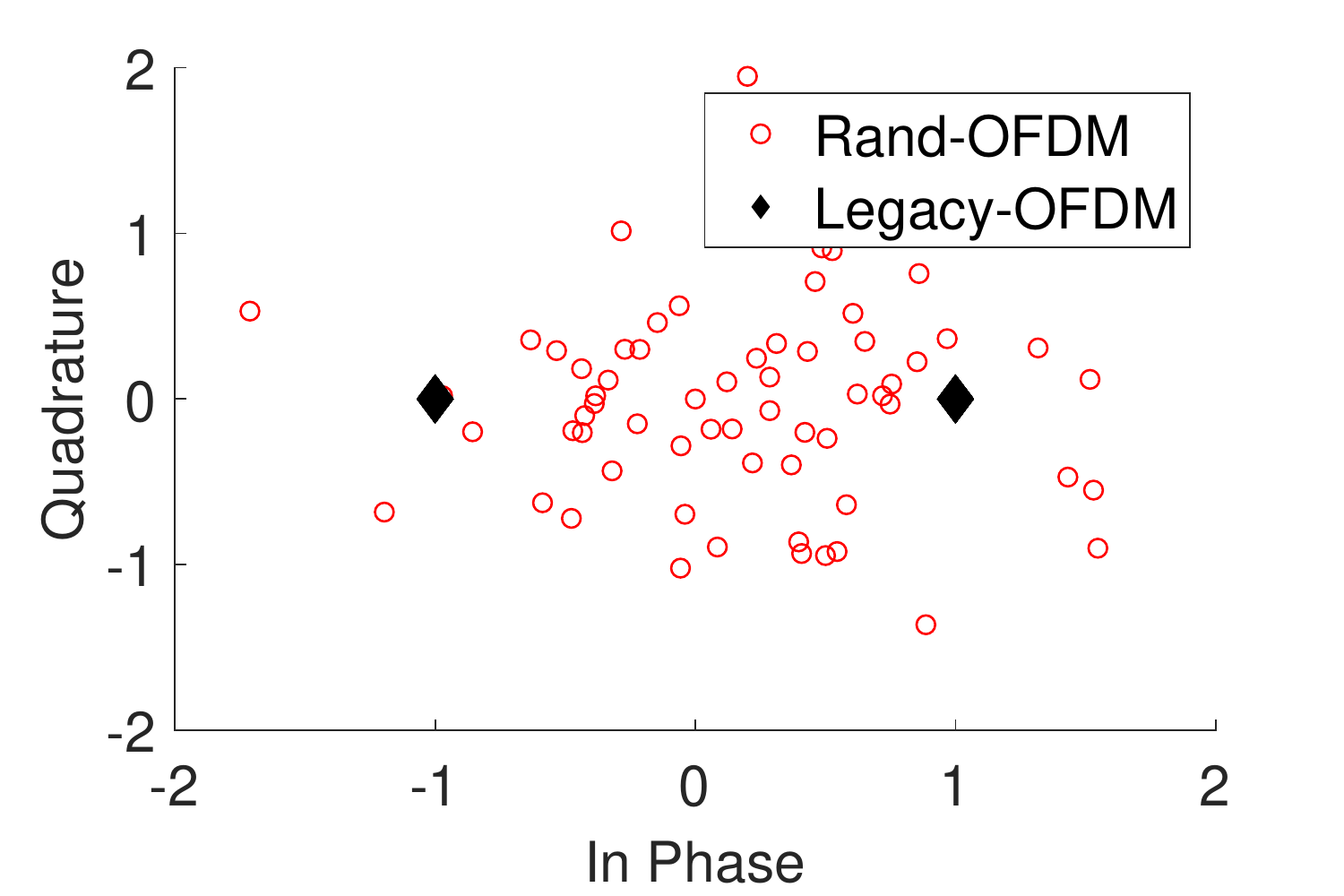}
        \caption{Constellation diagram of BPSK-modulated {\abbrev}.}
        \label{fig:rand_ofdm_const}
    \end{subfigure}
    \caption{Time domain and frequency domain representation of {\abbrev} at the transmitter.}
    \label{fig:randomized_out}
    \vspace{-10pt}
\end{figure*}

\noindent \underline{Key-based Time Domain Security}:
We designed a secret key based time domain modification of the OFDM signal to generate a \textit{new waveform}, {\abbrev}, which does not retain any OFDM properties that are essential to combat the multipath effects of the channel. At the receiver, we successfully reconstruct the signal by channel estimation and decryption of the {\abbrev} waveform.
\noindent \underline{Secured Training Signal}: We designed a novel training signal based on a shared phrase or data with the same randomization key, such that only the intended receiver is able to correctly decode it to estimate the channel accurately at the receiver. 

\noindent \underline{Clustering based Phase Offset Correction}: We scramble the signal in time domain, where the frequency domain pilots change in both phase and magnitude and cannot be used as known symbols for channel estimation or phase tracking. We introduced a \textit{unique} solution to track the residual phase offset by utilizing K-medoids clustering algorithm underneath.
    
\noindent \underline{Cryptanalysis}: We perform cryptographic analysis on the {\abbrev} signal, and provide insights on the resiliency of the waveform, specially for higher orders of FFT, as envisioned in next generation of wireless systems. 

\noindent \underline{Practical Evaluation}: In midst of mostly theoretical research in time domain physical layer security, we formulated the mathematical problem for {\abbrev} and provided a receiver design that is essential for successfully decoding the signal in multipath-rich environment. We also evaluated our system using extensive experiments over the air in an indoor environment for the protocol to be embraced for practical deployment.

\section{Related Work}
\label{sec:related}

\noindent
\textbf{Time domain security}:
A fairly decent amount of research has been done in recent years to modify the time domain signal to achieve security in physical layer. However, most of them are either theoretical or are dependent on impractical assumptions. In~\cite{infocom}, authors introduced an OFDM encryption scheme based on multiplying the real and the imaginary parts of the generated OFDM  by dynamic values based on a shared secret key between the transmitter and the receiver. This changes the signal values, which changes the Peak-to-Average Power Ratio (PAPR) of the OFDM symbol. The results are evaluated only for simulated AWGN channel, where there is no channel effects. Authors in~\cite{Post_IFFT} uses a random shuffling based on a secret key. The algorithm has been evaluated for a flat fading channel, where channel effects are negligible. These works do not attempt to introduce any channel or phase correction at the receiver due to theoretical nature of these research. 


In~\cite{cp}, authors propose a time domain physical layer security scheme based on modifying the length of the cyclic prefix (CP) to be equal to the channel impulse response of the intended user, whereas the channel impulse of \textit{Eve} may be longer than the intended user. This introduces inter symbol interference (ISI) to the received signal of the Eve. However, if \textit{Eve} has a better channel compared to \textit{Bob}, this technique fails.
In~\cite{129}, authors propose adding an artificial noise to the time domain OFDM signal such that when it passes through the receiver channel, it gets accumulated on the Cyclic Prefix. The receiver can decode the message after removing the CP, while the eavesdropper's signal can not be recovered due to the presence of the noise. This is also a theoretical attempt without any receiver modification in practical scenarios.


\noindent \textbf{Frequency domain secrecy}:
Modifying the frequency domain signal is a common technique to achieve secured~\cite{79, 83, ofdmind, Dummy} or covert communication~\cite{secret_radio_13}.
In~\cite{79}, the transmitter uses the non fading subcarriers to the intended user for data transmission. The assumption of this theoretical work is that \textit{Eve}'s channel has a completely different deep fading, which might not be a practical assumption.  
In~\cite{83}, the authors proposed an algorithm to provide pre-coder and post-coder matrices to make the channel matrix of the legitimate user to be diagonal. 
Also the authors in~\cite{ofdmind} proposed an optimal channel selection for channel indices to maximize the SINR for the legitimate user to achieve secure communication. 
In~\cite{Dummy}, the author proposed an encryption algorithm for OFDM system based on dummy data insertion of a portion of sub carrier to make performance degradation at the eavesdropper. It is to be noted that frequency domain modification retains the OFDM properties and reveals the waveform characteristics to the eavesdropper.

\noindent
\textbf{Space domain secrecy}: 
Use of MIMO antennas~\cite{101,103,104} to beamform towards the intended receiver and/or steer null towards the eavesdropper is another way to enable physical layer security. The two major assumptions in these set of work are a) \textit{Eve} will have fewer antenna elements, which cannot be accurate in the world of electronic warfare and b) \textit{Eve} is not in the path of beam pattern radiation, which is inaccurate specially when devices are getting smaller and can be placed anywhere in plain sight.

\section{Background}
\label{sec:back}

Orthogonal Frequency Division Multiplexing (OFDM) is a digital multicarrier modulation technique, where subcarriers are orthogonal to each other. This is achieved by Inverse Fast Fourier transform (IFFT) on the modulated data stream. For $N$ parallel data streams, $N-point$ IFFT is performed on the complex digital modulated signal $X(n)$. The time domain signal $x(n)$, can be expressed as:
\begin{equation}\label{eq:ofdm_tx}
    x[n] = IFFT(X[n]) = \sum_{i=1}^{N-1}{X(i)e^{\frac{j2\pi i n}{N}}} 
\end{equation}

In order for the IFFT/FFT to create an intersymbol interference (ISI) free channel, the channel must appear to provide a circular convolution. Hence, a cyclic prefix of length v is appended after the IFFT operation, resulting in N + v samples, which are then sent in serial through the wireless channel.

The duality between circular convolution in the time domain and simple multiplication in the frequency domain is a property unique to the Discrete Fourier Transform (DFT), which can be utilized to represent the received OFDM signal 
$Y(n) = H(n)X(n)$.

At the receiver, the cyclic prefix is discarded, and the N received symbols are demodulated, using an FFT operation, which results in N data symbols. The received frequency domain signal $Y(n)$ is given by:
\begin{equation}
    Y[n] = FFT[y(n)] = \sum_{i=1}^{N-1}{y(i)e^{\frac{-j2\pi i n}{N}}} 
\end{equation}
where $y$ is the received OFDM symbol in the time domain.

\section{System Design of {\abbrev}}
\label{sec:system}

\begin{figure*}
    \centering
    \includegraphics[width=0.75\linewidth]{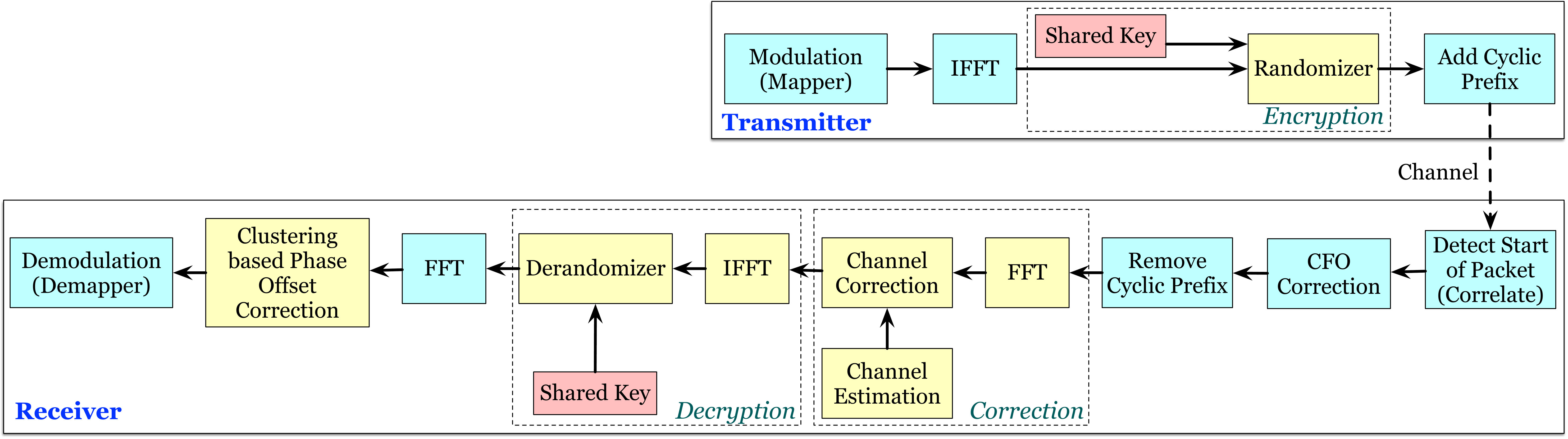}
    \caption{Transmitter and receiver blocks in {\abbrev}. Blocks in yellow are either added or modified in the legacy system.}
    \label{fig:system}
\end{figure*}

In this section, we introduce the system design of {\abbrev}, which predominantly has modifications at both the transmitter and the receiver side. The key idea of {\abbrev} lies in randomizing the FFT output or the time domain digital complex signals to loose the OFDM properties, such that it appears as a wideband noise to the eavesdropper. Figure~\ref{fig:randomized_out} shows such an example a) time domain signal after the IFFT output, followed by b) its randomized version, which is transmitted over the air. The c) frequency response of the transmitted {\abbrev} signal and d) its constellation plot show that we have successfully destroyed the OFDM properties for the signal to appear as a noise at the transmitter side, even before the channel impairments are introduced. Figure~\ref{fig:system} shows the transmitter and receiver modules, which are discussed in \S~\ref{sec:system_tx} and \S~\ref{sec:system_rx} respectively.

\subsection{Transmitter Design of {\abbrev}}
\label{sec:system_tx}


{\abbrev} transmitter is based on Wi-Fi~\cite{802_11_spec} like OFDM~\cite{ofdm} building blocks. 
We introduce a new block termed \textit{Randomizer} after the IFFT block in the OFDM transmitter chain, as shown in figure~\ref{fig:system}. The \textit{Randomizer} introduces time domain scrambling of the resultant time domain complex samples from the IFFT block. 
The time domain scrambling is based on a shared secret key between the transmitter and the receiver pair. In other words, this is the symmetric key that is used both in encryption and decryption process. For example, if the original OFDM symbol is $[a_0,a_1,a_2,a_3,a_4,a_5]$, then the resulted randomized sequence is $[a_5,a_3,a_1,a_2,a_4]$, where the key is $[5,3,1,2,4]$. The cyclic prefix (CP) is added after that to be able to extract frequency response of the channel at the receiver, which is detailed in \S~\ref{sec:system_rx}.

As the \textit{Randomizer} block is added in the OFDM transmitter chain, the time domain {\abbrev} signal, $x_{t}$, can be expressed as:
\begin{equation}\label{eq:randTx}
    x_{t} = P_{CP}RF^{-1}X_{F}
\end{equation}
where $P_{CP}$ is the cyclic prefix matrix, $R$ is the randomizer matrix and $F^{-1}$ is the inverse Fourier Transform Matrix.

By randomizing in time domain, the secured transmitted wave form has lost all the OFDM properties. In other words, the {\abbrev} transmitted wave form $X_{t}$ can given by:
\begin{equation}
    X_{t} = FT x_{t} = FR F^{-1}X_{F}
\end{equation}
 where $T$ is the truncation matrix for cyclic prefix removal and $F$ is the N-FFT matrix and can be given as:
 \begin{equation}
  F=  \begin{bmatrix}
         W^{0,0} & W^{0,1} & .........&W^{0,N-1}\\
         W^{1,0} & W^{1,1} & .........&W^{1,N-1}\\
         .\\
         .\\
         W^{N-1,0} & W^{N-1,1} & .........&W^{N-1,N-1}
    \end{bmatrix}
\end{equation}
 where $W^{n,k}=e^{-2j\pi\frac{nk}{N}}$ and  N is the FFT size. 
\begin{equation}\label{eq:randTD}
 F R F^{-1} =\frac{1}{N} \sum{e^{2j\pi\frac{n(k-m)}{N}}}= \begin{cases}
 I,&R=I \\
 Q,&R \ne I
 \end{cases}
\end{equation}
where $I$ is the identity matrix and $Q$ is the generated transformation matrix due to the presence of $R$.
 
From equation~\ref{eq:randTD}, we can conclude that if there is no randomization in the time domain samples (i.e $R_F = I$), the OFDM symbol retains its orthogonality property. Otherwise, a magnitude and phase noise is added to the received OFDM symbol. This distorts the original OFDM symbol by transforming the original constellation into a random constellation pattern which depends on the randomization matrix $R$. We would like to highlight here that the changes introduced in this stage do not depend on the original modulation order of the transmitted constellation. Moreover, the transformation matrix spreads the power of the OFDM symbol over the whole bandwidth including the guard bands. 
Figure~\ref{fig:randomized_out} shows the transformation of an OFDM symbol to a {\abbrev} symbol. In this case, 52 subcarriers were modulated with BPSK modulated signal, generating the time domain OFDM signal as in figure~\ref{fig:ofdm}. The resultant signal goes through a \textit{Randomizer} to generate the {\abbrev} waveform as shown in figure~\ref{fig:rand_ofdm}. If we perform FFT on this waveform, we notice that the magnitude of the subcarriers are varying randomly, as well as the power spreads to all 64 subcarriers and not confined to only 52, as we started with. This is evident in figure~\ref{fig:rand_ofdm_fft}, which is due to the matrix multiplication of $R$. Figure~\ref{fig:rand_ofdm_const} shows the phase noise that is induced due to the randomization. These imperfections require significant modifications in the receiver design, which is explained in \S~\ref{sec:system_rx}.

\begin{figure}
    \centering
    \includegraphics[width=0.9\linewidth]{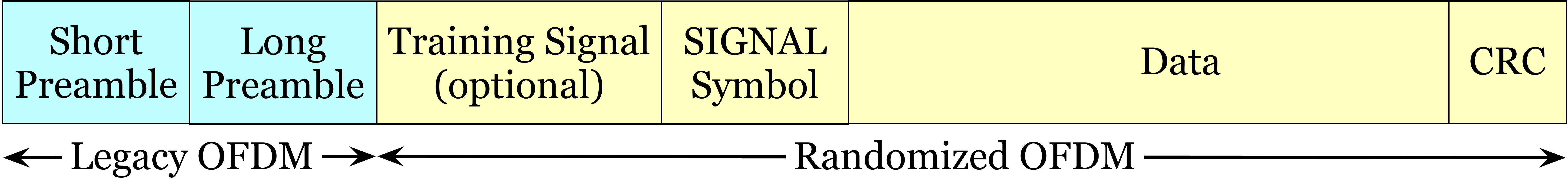}
    \caption{Packet Structure of {\abbrev}.}
    \label{fig:packet}
\end{figure}

The packet structure of {\abbrev} is shown in figure~\ref{fig:packet}, where the short and long preambles of legacy 802.11~\cite{802_11_spec} system are used to detect the start of the packet. Rest of the MAC layer packet structure remains same except they are modified just before transmission to the {\abbrev} waveform. A training signal of one OFDM symbol duration is appended to improve the channel estimation at the intended receiver, the design of which is detailed in \S~\ref{subsec:channel}.

\subsection{Receiver Design of {\abbrev}}
\label{sec:system_rx}

The receiver design of {\abbrev} is based on legacy 802.11a/g~\cite{802_11_spec} receiver blocks, where yellow colored blocks are the ones which we added or modified, as shown in figure~\ref{fig:system}. Receiver design starts with a packet detection block, followed by Carrier Frequency Offset (CFO) correction and removal of cyclic prefix. FFT follows right after that to convert the data to the frequency domain. Based on the modifications at the transmitter, it might seem that we need to insert the \textit{Derandomizer} block just before the data goes into the FFT block. One of the major components before the FFT is the channel correction, as the received waveform can be represented as: 
\begin{equation}\label{eq:rand_rx}
    Y_{t} = H_{t}X_{t} + N_{t} = H_{t}P_{CP}RF^{-1}X_{F} + N_{t}
\end{equation}
where $H_{t}$ is the channel impulse response in time domain and $N_{t}$ is the noise. At this point, we need to estimate and correct the channel impairments at the receiver. Time domain channel estimation and correction has been shown to be computationally more expensive~\cite{time_dom_1}, 
due to which we choose to perform it in frequency domain. Derandomizing the signal before extracting the channel information would make the channel estimation problem intractable. Hence, we convert the signal to frequency domain to extract the frequency domain channel response, which is first corrected on the signal.
Once the cyclic prefix is removed, we perform N-point FFT to convert the time domain signal $Y_{t}$ to frequency domain. The received waveform in frequency domain can then be expressed as:
\begin{equation}\label{eq:rx_freq}
    X_{rF} = FTY_{t} = H_{F}FR F^{-1}X_{F} + N_{F}
\end{equation}
where $H_{F}$ is the channel frequency response. Channel estimation for {\abbrev} is discussed in details in \S~\ref{subsec:channel}. In the following discussion, we assume that we know the channel $H_{F}$ and inverse it yielding received frequency domain waveform $\Tilde{X}_{rF}$ without channel impairments.
\begin{equation}
    \Tilde{X}_{rF} = H_{F}^{-1} X_{rF} = F R F^{-1}X_{F} + \Tilde{N}_{F}
\end{equation}
This is the output of the channel correction block as shown in figure~\ref{fig:system}, which is in frequency domain. We introduced the randomization in time domain, and hence it is necessary to convert it back to time domain as a part of the decryption block. After the signal is passed through the IFFT block of the decryption process, it can be represented as
\begin{equation}
    F^{-1}\Tilde{X}_{rF} = F^{-1}FR F^{-1}X_{F} + \Tilde{N}_{F} = R F^{-1}X_{F} + \Tilde{N}_{F}
    \label{eq:WF_A_CH_Est}
\end{equation}
Now, this time domain signal is passed through the \textit{Derandomization} block to yield the correct time domain data symbols.
\begin{equation}\label{eq:decrypted1}
    R^{-1}R F^{-1}X_{F} + \Tilde{N_{F}} = F^{-1}X_{F} + \Tilde{N}_{F}
\end{equation}
The final step is to perform FFT on equation~\ref{eq:decrypted1} to demodulate the data subcarriers.
\begin{equation}\label{eq:decrypted}
    FF^{-1}X_{F} + \Tilde{N_{F}}=X_{F} + \Tilde{N}_{F}
\end{equation}
We introduce another Phase Offset Correction phase, which is detailed in \S~\ref{subsec:cluster}.

\section{Distortions from Channel and Hardware}
\label{sec:improve}


 \begin{figure*}
    \centering
    \begin{subfigure}[b]{0.23\textwidth}
        \includegraphics[width=\textwidth]{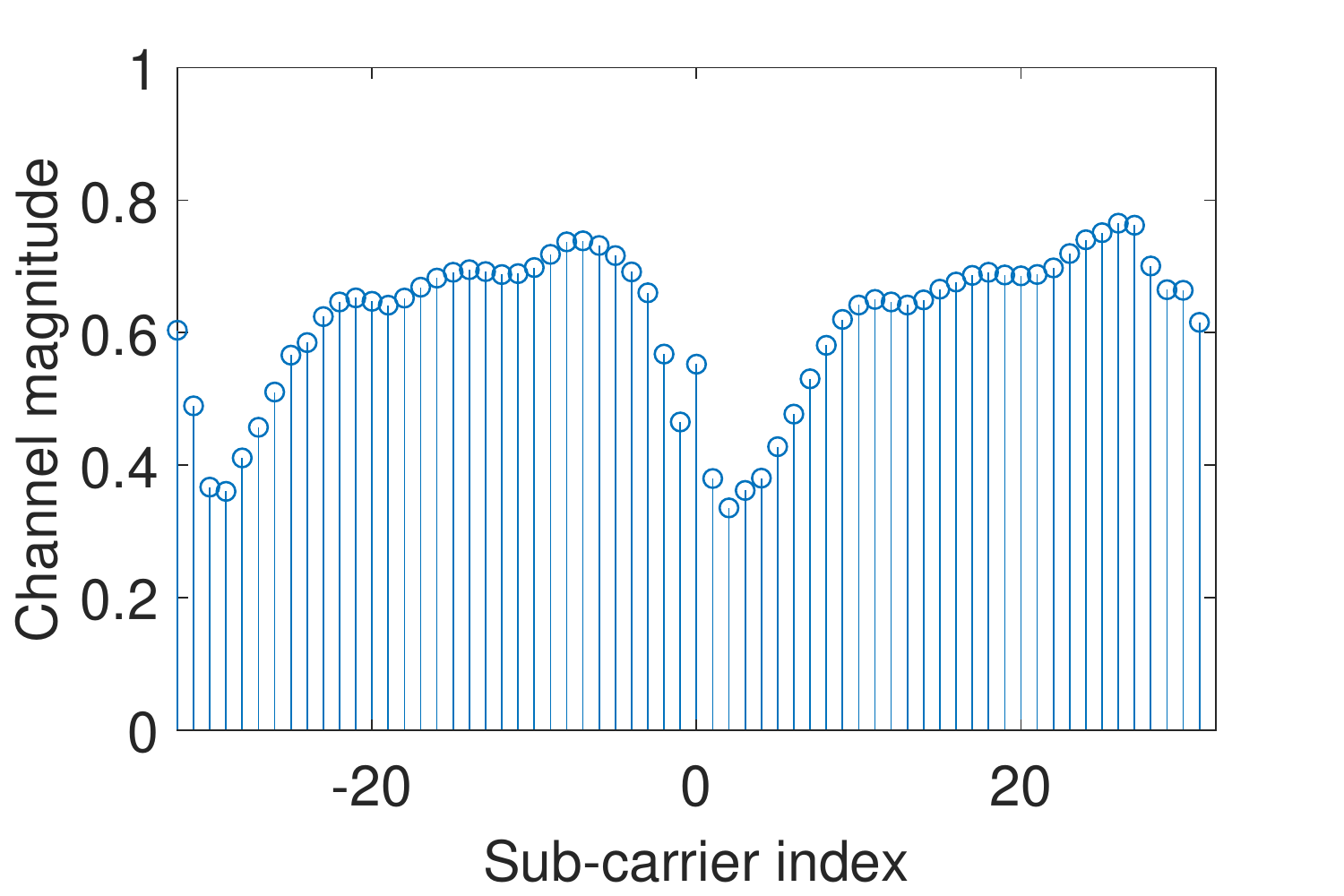}
        \caption{Actual Channel (Mag)}
        \label{fig:channel_m}
    \end{subfigure}
    \quad
    \begin{subfigure}[b]{0.23\textwidth}
        \includegraphics[width=\textwidth]{Figures/Channel_mag.pdf}
        \caption{Estimated Channel (Mag)}
        \label{fig:ch_training_m}
    \end{subfigure}
    \quad
    \begin{subfigure}[b]{0.23\textwidth}
        \includegraphics[width=\textwidth]{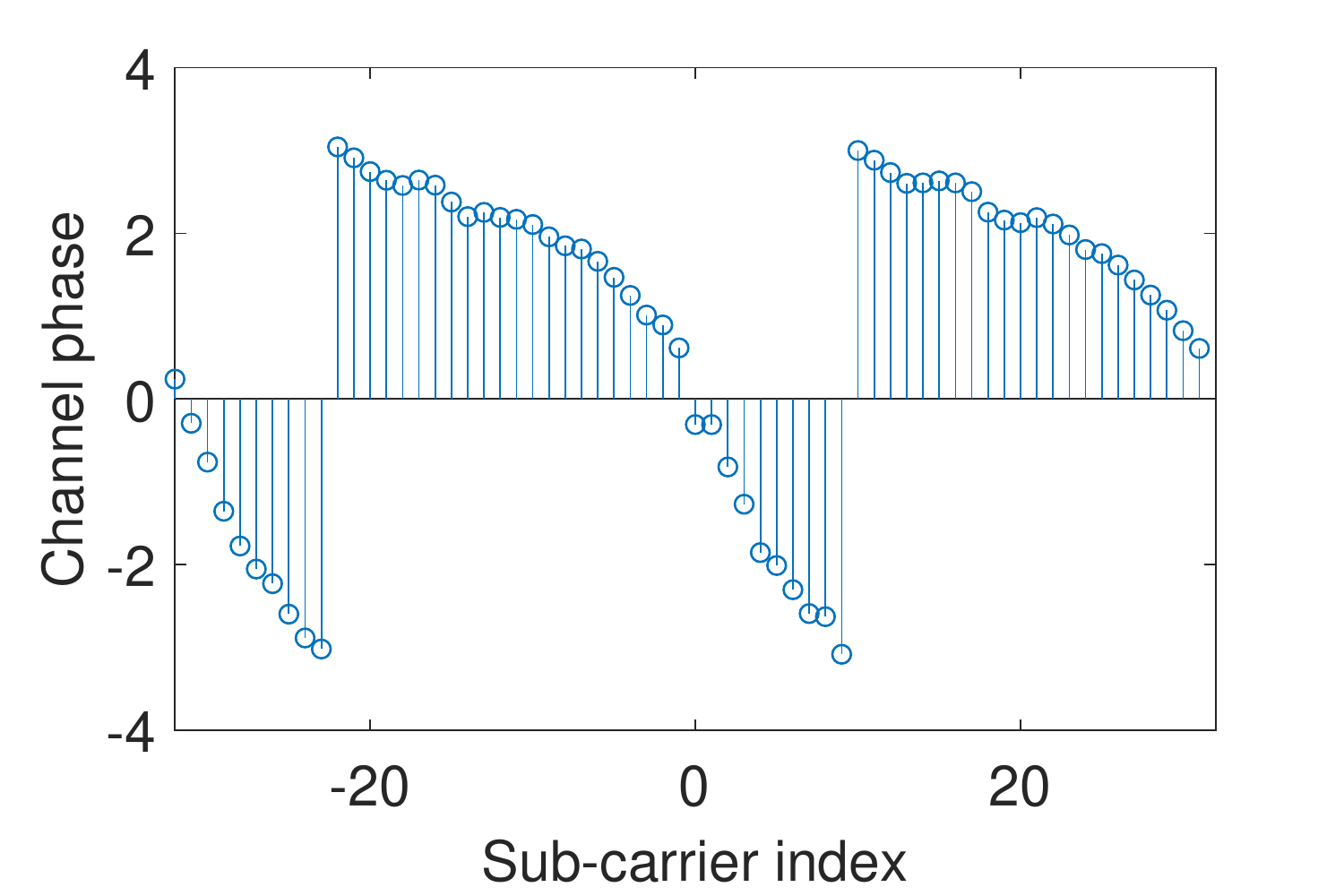}
        \caption{Actual Channel (Ph)}
        \label{fig:channel_p}
    \end{subfigure}
    \quad
    \begin{subfigure}[b]{0.23\textwidth}
        \includegraphics[width=\textwidth]{Figures/Channel_ph.pdf}
        \caption{Estimated Channel (Ph)}
        \label{fig:ch_training_p}
    \end{subfigure}
    \caption{Channel Estimation (Magnitude \& Phase) of frequency selective channel for training signal based estimation technique.} 
    \label{fig:channel_est}
  \vspace{-10pt}
\end{figure*}


Imperfections at the received are induced due to the wireless multipath channel and the underlying hardware. 
In this section, we illustrate the methodologies that we introduce to estimate and counteract the imperfections to yield better performance.

\subsection{Training signal based channel estimation}
\label{subsec:channel}

Accurate estimation of channel is essential in high frequency selective fading channels, specially at higher order modulations, where minimal error in channel estimation will lead to significant error in demodulation. In \S~\ref{sec:system_rx}, we assumed complete knowledge of the channel $H_{F}$, which is an incorrect assumption in practical systems. Partial channel estimation on only 52 subcarriers of long preamble and extrapolating that for 64 subcarriers, as required for {\abbrev}, yields poor performance in multipath channels.
Hence, we design a training signal that spans all 64 subcarriers to better estimate the channel at the intended receiver. We use a shared secret data and randomize it based on the same shared key between \textit{Alice} and \textit{Bob} to create one BPSK modulated {\abbrev} signal, as described in \S~\ref{sec:system_tx}. This OFDM symbol is transmitted after the preamble as shown in figure~\ref{fig:packet}. Channel estimation at the receiver requires knowledge of transmitted waveform, which in this case is secured by the shared key. Hence, we add another layer of security, where Eve will not be able to estimate the channel accurately as the training signal is dependent on the shared secret key, and physical layer authentication can be initiated using the training signal \cite{Auth1,Auth2}.

If $X_{Tr}$ is the BPSK-modulated OFDM signal, then the {\abbrev} training signal can be derived from equation~\ref{eq:randTx} as $x_{Tr} =P_{CP}RF^{-1}X_{Tr}$. The received training signal can be derived from equation~\ref{eq:rand_rx} as 
\begin{equation}
y_{Tr} =H_{t}P_{CP}RF^{-1}X_{Tr}+N_{Tr}
\end{equation}
Since $X_{Tr}$ and $R$ matrices are shared between the transmitter and the intended receiver only, the channel can be estimated by only \textit{Bob}. Ignoring the noise, channel frequency response can be estimated by:
\begin{equation}
    H_{F}= \textstyle \frac{FTy_{Tr}}{X_{Tr}} 
\end{equation}
Figure~\ref{fig:channel_est} shows the actual and the estimated phase and magnitude of the channel in frequency selective fading conditions. It is obvious that training signal channel estimation gives an accurate channel estimation for the whole 64 sub-carriers rather than the preamble channel estimation which only concerns on 52 sub-carriers .

\begin{figure}
    \centering
    \begin{subfigure}[b]{0.23\textwidth}
        \includegraphics[width=\textwidth]{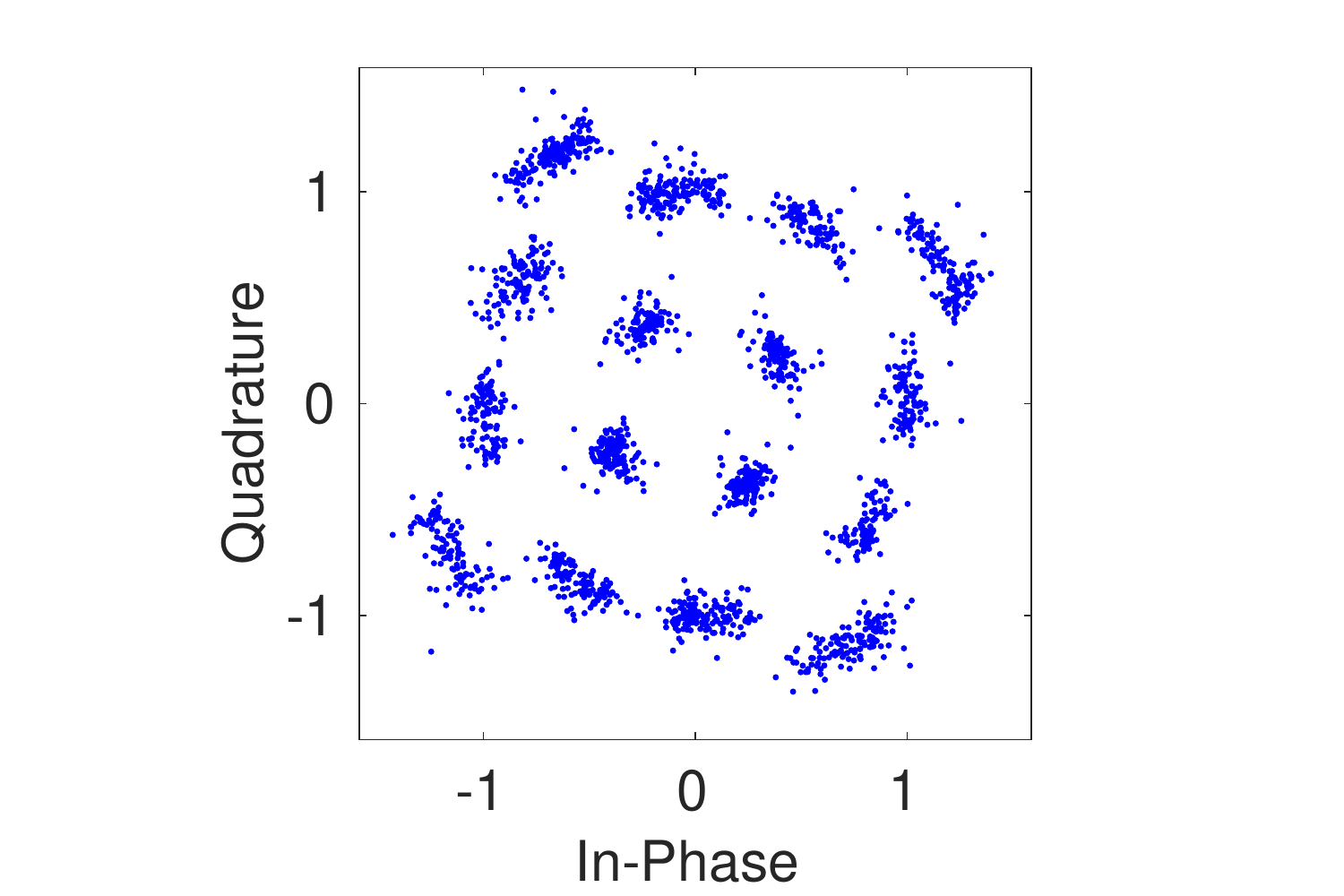}
        \caption{Before Phase correction}
        \label{fig:16qam_before}
    \end{subfigure}
    \quad
    \begin{subfigure}[b]{0.23\textwidth}
        \includegraphics[width=\textwidth]{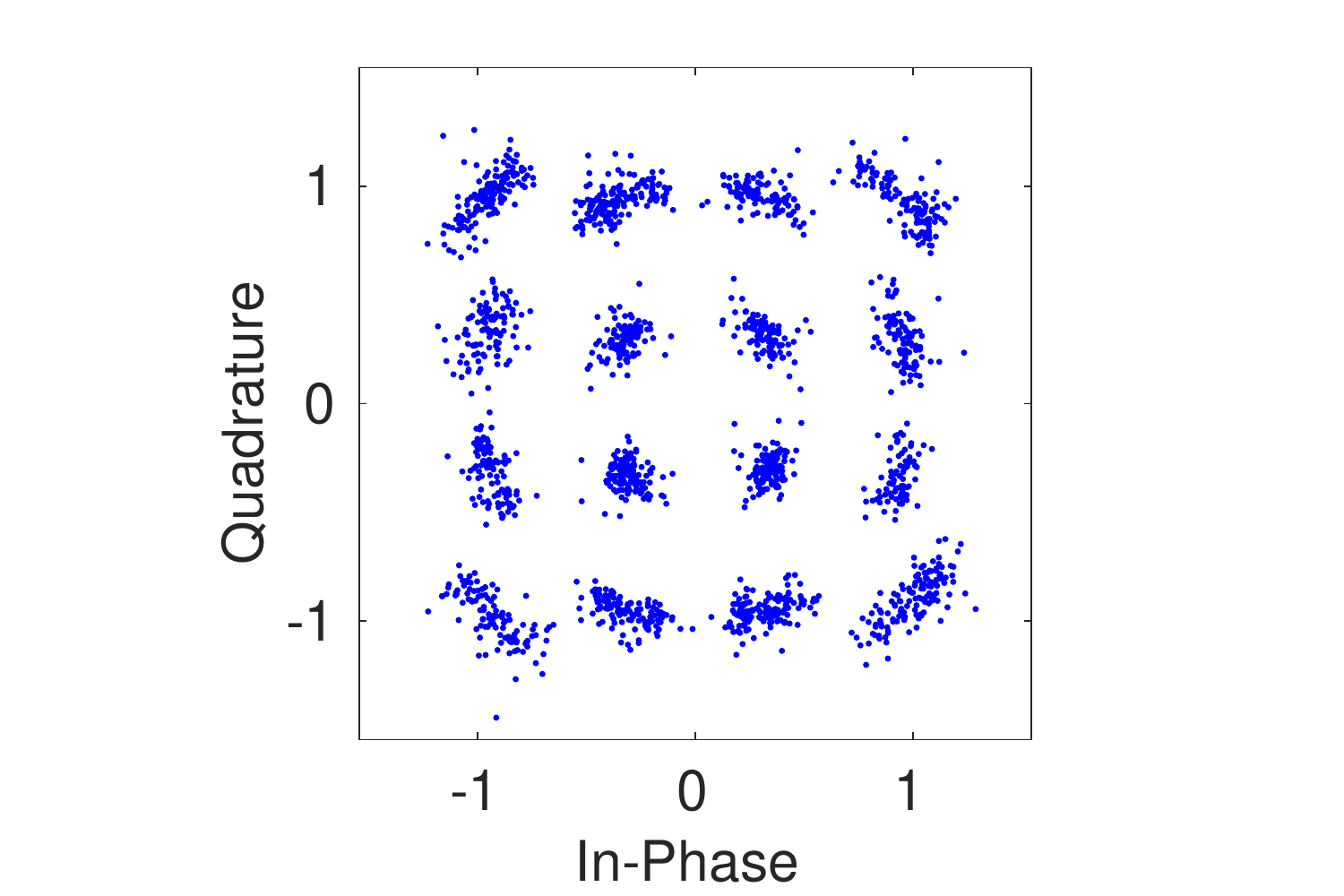}
        \caption{After Phase correction}
        \label{fig:16qam_after}
    \end{subfigure}
    \caption{Clustering based residual phase offset correction for over the air indoor experiments using 16QAM modulation.}
    \label{fig:phase_corr}
\end{figure}

\subsection{Clustering based phase offset correction}
\label{subsec:cluster}

Device impairments introduce difference between the carrier frequency of the receiver and that of the transmitter. Coarse and fine carrier frequency offset correction blocks are introduced at the receiver to estimate and correct those offsets based on short and long preambles respectively, as shown in figure~\ref{fig:system}. Residual carrier frequency offset of conventional OFDM received waveform is calculated based on the pilot subcarriers. In {\abbrev}, the pilots inserted in frequency domain do not capture the channel frequency response at those subcarriers as the pilot energy gets spread due to the process of randomization. Hence, we introduce a clustering based algorithm to track the residual phase offset due to hardware impairments. It is to be noted here that the purpose of this block is not to estimate the variation in channel per OFDM symbol, which can be accurately estimated by inserting pilot subcarriers in every OFDM symbol. This block is intended to estimate and correct the offset between the transmitter and receiver pairs, which do not change from one OFDM symbol to the next.

We use K-mediods clustering algorithm~\cite{kaufman1987clustering}, where the input to the algorithm is the In-Phase and Quadrature values of the constellation points of all the subcarriers of all received OFDM symbols in a packet and the number of expected clusters based on the modulation order. The resultant $C$ cluster centers are then examined to estimate the residual phase offset. 
We choose the farthest cluster points in each quadrant to determine the phase offset. There are also the highest energy point indicating a total of 4 cluster centers for QAM or QPSK modulations and only 2 for BPSK. The rationale for choosing the highest energy point within a quadrant is that there exists only one such point in the transmitted constellation to which it can be mapped to. This is a generic approach and can be scaled to even higher order modulations, like 256-QAM and beyond.
The phase offset is then calculated per quadrant as:
\begin{equation}
    \theta_{estimated,i}=arg(X_{Ti}/C_{max,i})
\end{equation}
where $X_{Ti}$ is the farthest transmitted constellation point within a quadrant and $C_{max,i}$ is the maximum energy cluster center of the same quadrant $i$. Averaging 2 values in BPSK or 4 for other modulation orders, we calculate residual phase offset as: 
\begin{equation}
    \theta_{estimated}= \textstyle \frac{1}{M}\sum_{i=1}^{M} \theta_{estimated,i}
\end{equation}
where $M$ is 2 for BPSK and 4 for other modulations.
In the last step, we correct the residual phase offset by multiplying the estimated phase to the received signal to generated the corrected signal, $X_{Fc}$, which can then be used for demodulation.
\begin{equation}
    X_{Fc}=X_{F}e^{j\theta_{estimated}}
\end{equation}
where $X_{F}$ is the received OFDM signal before correction. Figure~\ref{fig:phase_corr} shows an example scenario of over-the-air experiments at 15dB SNR, where the residual phase offset is corrected based on the proposed clustering algorithm.

\section{Cryptanalysis}
\label{sec:cryptanalysis}
In this section, we perform security analysis on {\abbrev} to evaluate its resiliency against various types of attacks. For simplicity, let's assume that the received OFDM symbol at \textit{Eve} is:
\begin{equation}
    Y_{E} = FRF^{-1}X_{F}
    \label{Rec_eq}
\end{equation}
According to Shannon secrecy~\cite{shannon1949communication}, the system can be perfectly secure if the key size equals to the data size such that:
  \begin{equation}
    E(R_{i}) \geq E(X_{F})
\end{equation}
where $E(X)$ is the entropy of the random variable $X$.
This analysis can easily be realized in bit level, however in physical layer we observe it from two different viewpoints. \textit{First}, the system achieves perfect secrecy on symbol level as both the data symbol size and the key size are equal to the FFT size N. In other words, \textit{Eve} can not deduce any information with one symbol.
\textit{Second}, if the system has a set of keys $\textit{K} = [K_{1}, K_{2}, K_{3}, ..., K_{V}]$, and OFDM frame $\textit{X}=[X_{1}, X_{2}, X_{3}, ..., X_{n}]$, the mutual information between the encrypted and original symbols equal to:
\begin{equation}
\lim_{V\to n} I(X_{E},X_{F})=0    
\end{equation}

\subsection{Brute-force attack}
In this attack, the eavesdropper knows the encryption and the decryption algorithm, however the key is unknown. The eavesdropper performs an exhaustive search on all possible keys to decrypt the cipher data (i.e the encrypted data). The strength of the algorithm is proportional to the length of the key, which is the size of FFT for {\abbrev}. The number of all possible keys $L$ is given by $L=N!$, where $N$ is the FFT size. So the probability of success $P_{s}$ to predict the key is uniformly distributed and is given by $P_{s} = \frac{1}{L}$. FFT size has been chosen to be 64, which is minimum for 802.11a/g~\cite{802_11_spec}, and can be a much larger value of 1024 or 2048 in newer wideband Wi-Fi and 5G standards.

\subsection{Cipher text attack}
In this attack, $Y_{E}$ is only available with the decryption function $D$. \textit{Eve} tries to predict $R_{E}$ by attempting different keys based on the statistical properties on the cipher data. However, since the security layer is introduced in physical layer, especially in time domain, the statistical properties of the received waveform does not change, for example received power or the peak to average power ration (PAPR). So \textit{Eve} has to try all possible keys to decrypt the data, resulting in Brute-force attack.
    
\subsection{Chosen plain text attack}
In this attack, $Y_{E}$ and $X_{F}$ are known at the eavesdropper. In other words, \textit{Eve} knows some pattern and the corresponding cipher patterns. For fixed key, \textit{Eve} can solve equation~\ref{Rec_eq} and deduce the corresponding $R_{E}$. 
However the security of the algorithm can be increased by using a defined shared set of keys $K=[K_{1}, K_{2}, ..., K_{V}]$, and the selected shared key changed dynamically from one symbol to another based on certain distribution $f_{K}(K)$. In this case, \textit{Eve} has to deduce the statistical properties of the key distribution for multiple attacks assuming $K$ is known.

\section{Evaluation}
\label{sec:eval}

In this section, we present the performance analysis of the {\abbrev} system in comparison with legacy 802.11a/g in different channel models. Although we do not use any pilot symbols, they are still inserted as in the legacy system such that same number of data bits are transmitted for both the cases. We used MATLAB to encrypt and decrypt the signals and used the channel models to perform extensive simulations for both legacy OFDM and {\abbrev}. For the rest of the paper, the suffixes (L) and (R) are used for legacy OFDM~\cite{802_11_spec} and {\abbrev} transmissions with full channel knowledge respectively. In addition, we use (R-T) using training signal based channel estimation.

\begin{table}
  \centering
  \begin{tabular}{lccc}
    \hline
    Channel & AWGN & Flat & Frequency Selective \\ \hline
    Model & - & Rayleigh & Rayleigh \\
    No. of taps & 0 & 1 &6 \\ 
    Path delays & 0 & 0 & [0,100,200,300,500,700] ns\\ 
    Path loss & 0 & 0 & [0,-3.6,-7.2,-10.8,-18,-25.2] dB\\
    Doppler & 0 & 0 & 3Hz \\ \hline
  \end{tabular}
  \captionof{table}{Channel Models} 
  \label{tbl:channel}
\end{table}

\subsection{Without Channel Estimation}
We evaluate the performance of {\abbrev} receiver blocks as described in \S~\ref{sec:system_rx} in the basic scenario, where no channel estimation is required.

\begin{figure*}
    \centering
    \begin{subfigure}[b]{0.245\textwidth}
        \includegraphics[width=\textwidth]{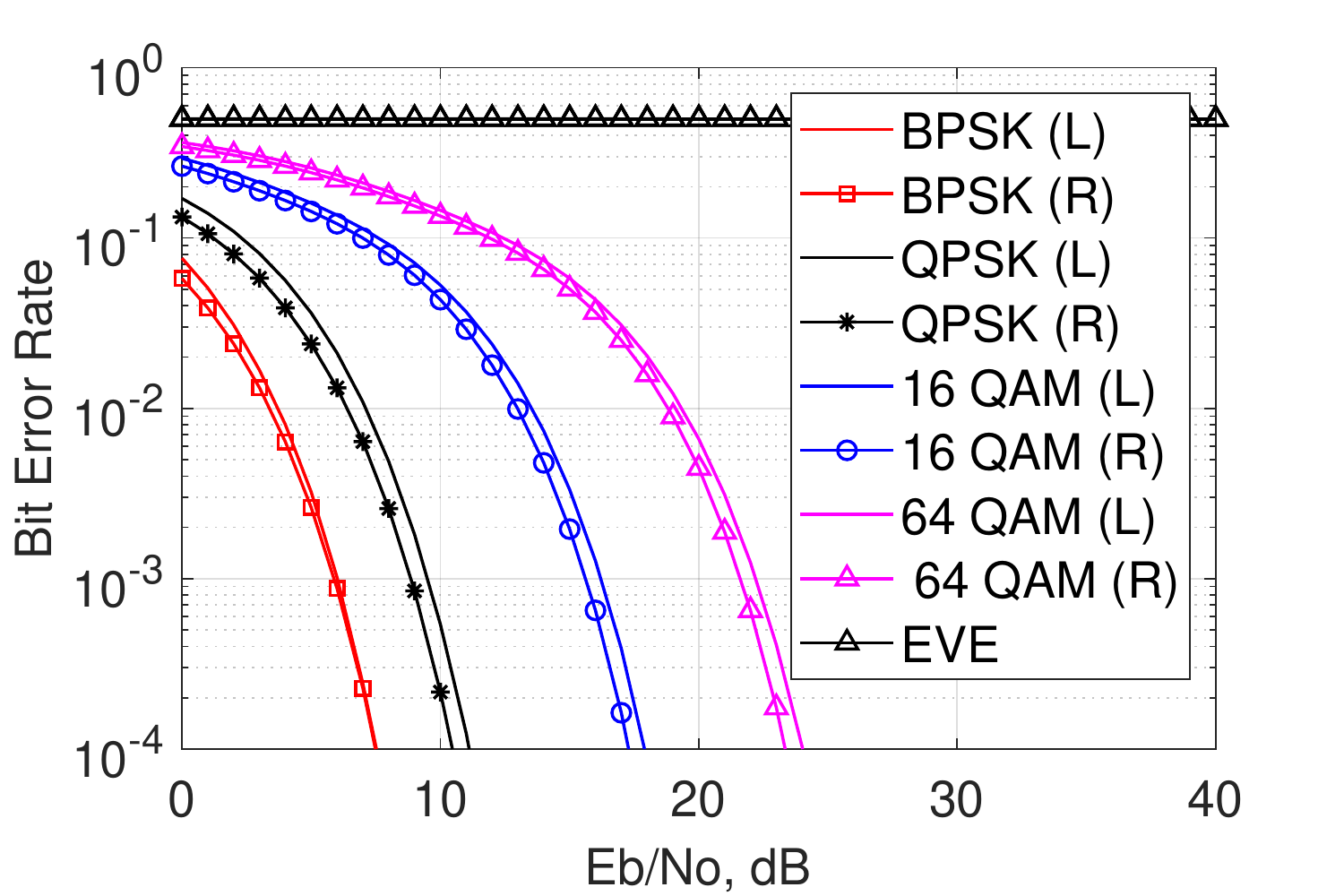}
        \caption{AWGN Channel.}
        \label{BER_AWGN_Trad}
    \end{subfigure}
    \begin{subfigure}[b]{0.245\textwidth}
        \includegraphics[width=\textwidth]{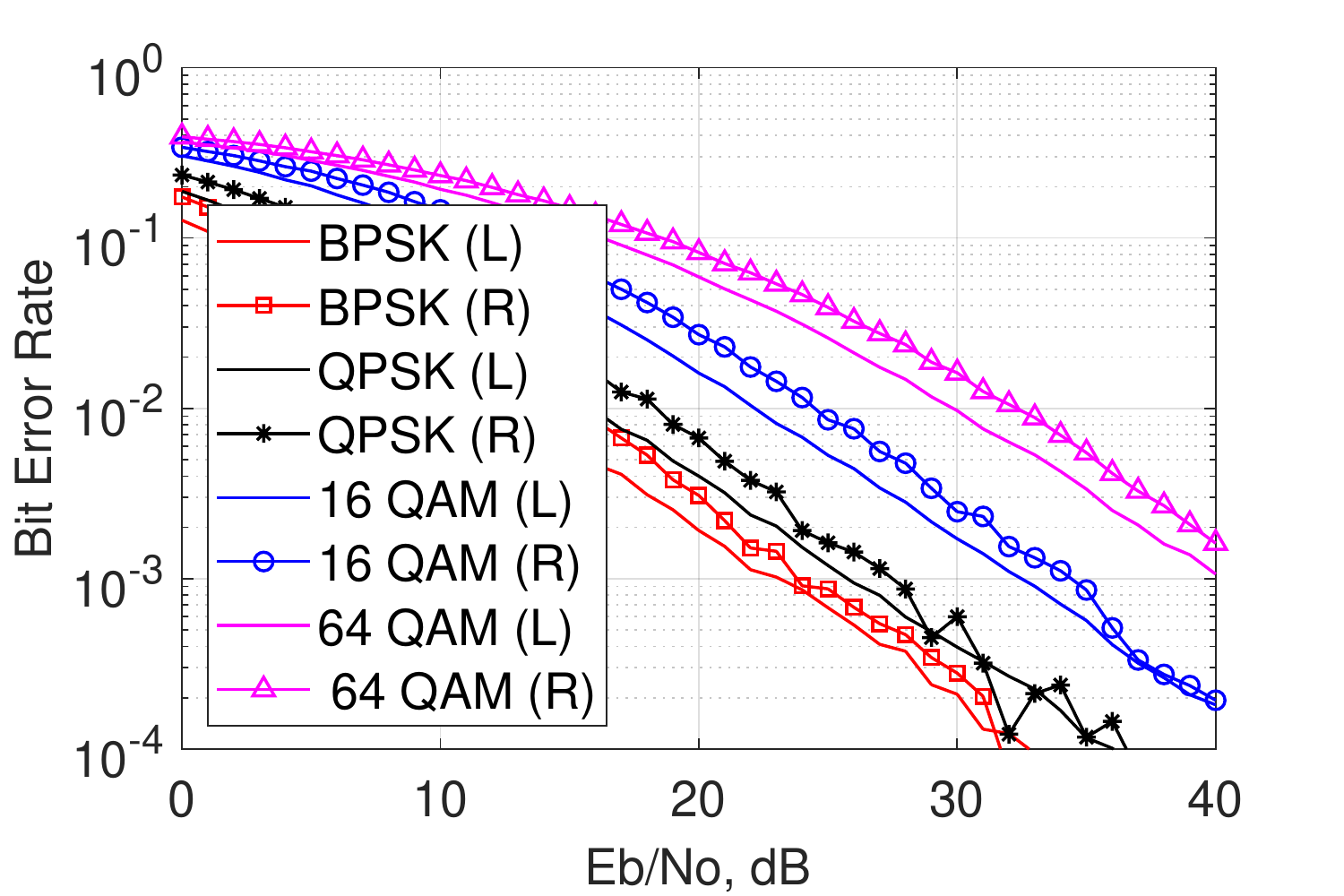}
        \caption{Indoor channel.}
        \label{BER_Full_CSI}
    \end{subfigure}    
    \begin{subfigure}[b]{0.245\textwidth}
        \includegraphics[width=\textwidth]{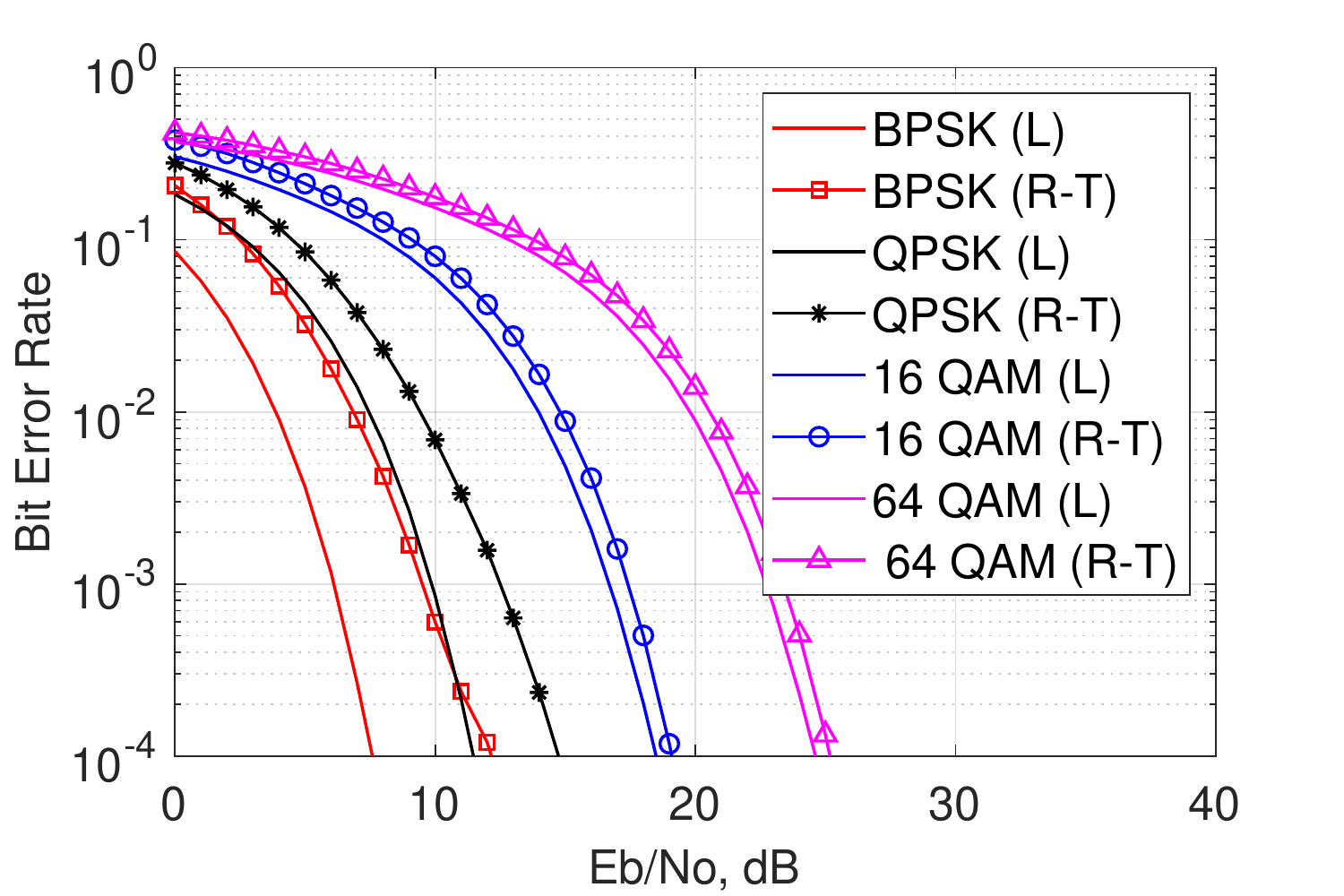}
        \caption{Flat fading channel.}
        \label{fig:train_flat}
    \end{subfigure}
    \begin{subfigure}[b]{0.245\textwidth}
        \includegraphics[width=\textwidth]{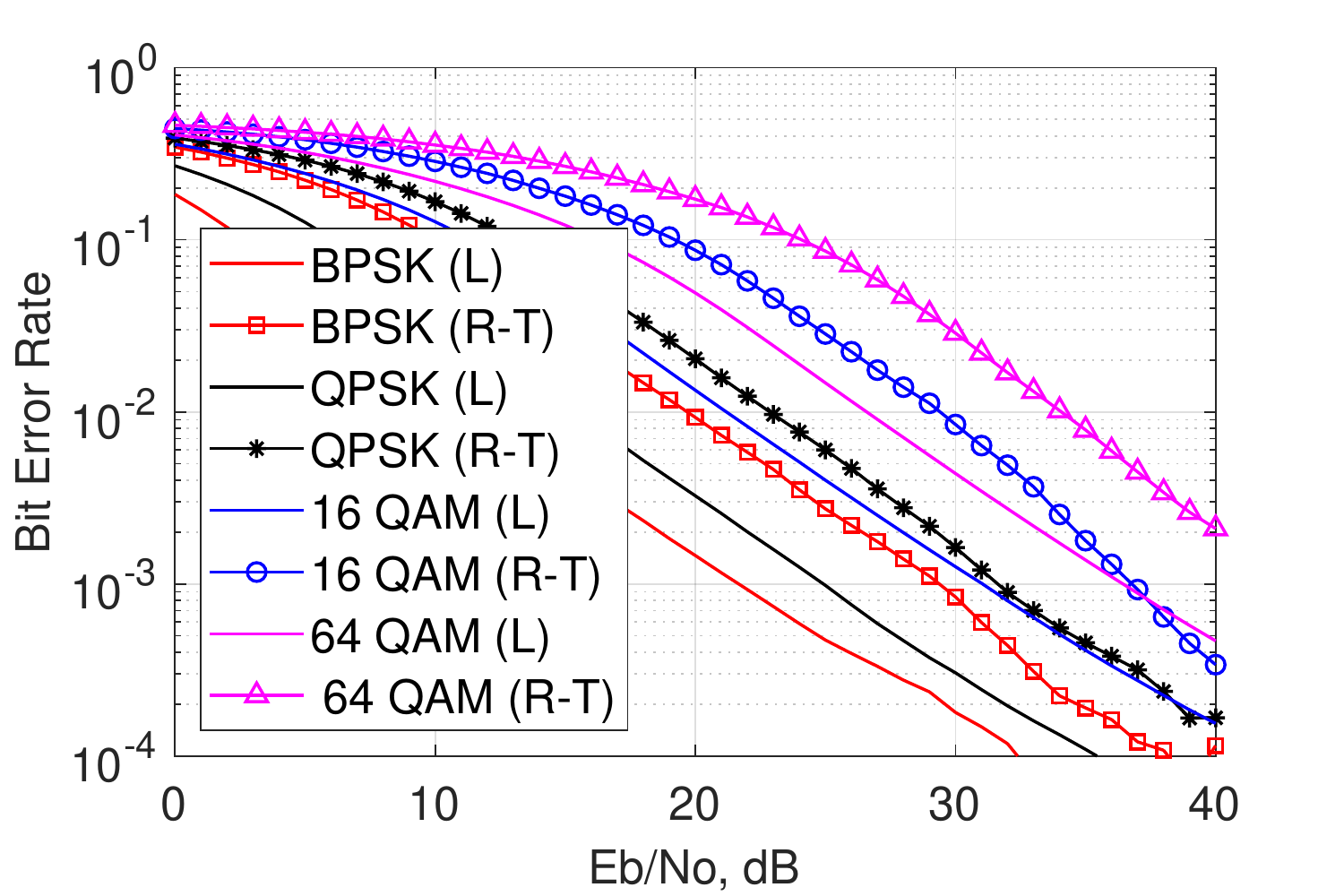}
        \caption{Indoor channel.}
        \label{fig:train_sel}
    \end{subfigure}    
    \caption{BER of {\abbrev} compared to legacy OFDM. Figures (a,b) with complete channel knowledge and (c,d) with training signal channel estimation.} 
    \label{fig:ber_preamble}
  \vspace{-15pt}
\end{figure*}

\subsubsection{AWGN Channel}
The first case for evaluation is Additive White Gaussian Noise (AWGN) channel, where the channel frequency response matrix $H_{F}=I$, since both time and frequency coefficients are unity. Figure~\ref{BER_AWGN_Trad} shows the BER performance of {\abbrev} in AWGN channel for different modulation orders. Performance of {\abbrev} is close to that of legacy OFDM signal. This is due to the fact that OFDM structure, even when lost, can be reconstructed back at the receiver in the absence of channel effects. From equation~\ref{eq:WF_A_CH_Est}, it is evident that if the eavesdropper does not have the key to generate the matrix $R$, she can not decode the packet. This is shown in the results as well, where the BER curve for \textit{Eve} remains constant and never go down even at higher SNRs. This is a major advantage of using key-based physical layer security over channel based encryption techniques. 
We choose not to show the BER curve for \textit{Eve} as she was unable to decode in the best possible channel.

\subsubsection{Complete channel knowledge}

In this section, the performance of the proposed system is evaluated assuming that we have full channel knowledge (i.e $H_{F}$ is known). Based on ITU-R recommendation, we assume frequency selective indoor channel model with parameters shown in table~\ref{tbl:channel}. We assume that legacy receiver also knows the $H_{F}$ matrix. Figure~\ref{BER_Full_CSI} shows the performance of {\abbrev} in a frequency selective channel when the channel is known at the receiver. If channel is known, the SNR gap between legacy OFDM and {\abbrev} is due to the modification of the waveform, which cannot be retrieved at the receiver.


\begin{figure*}
    \centering
    \begin{subfigure}[b]{0.23\textwidth}
        \includegraphics[width=\textwidth]{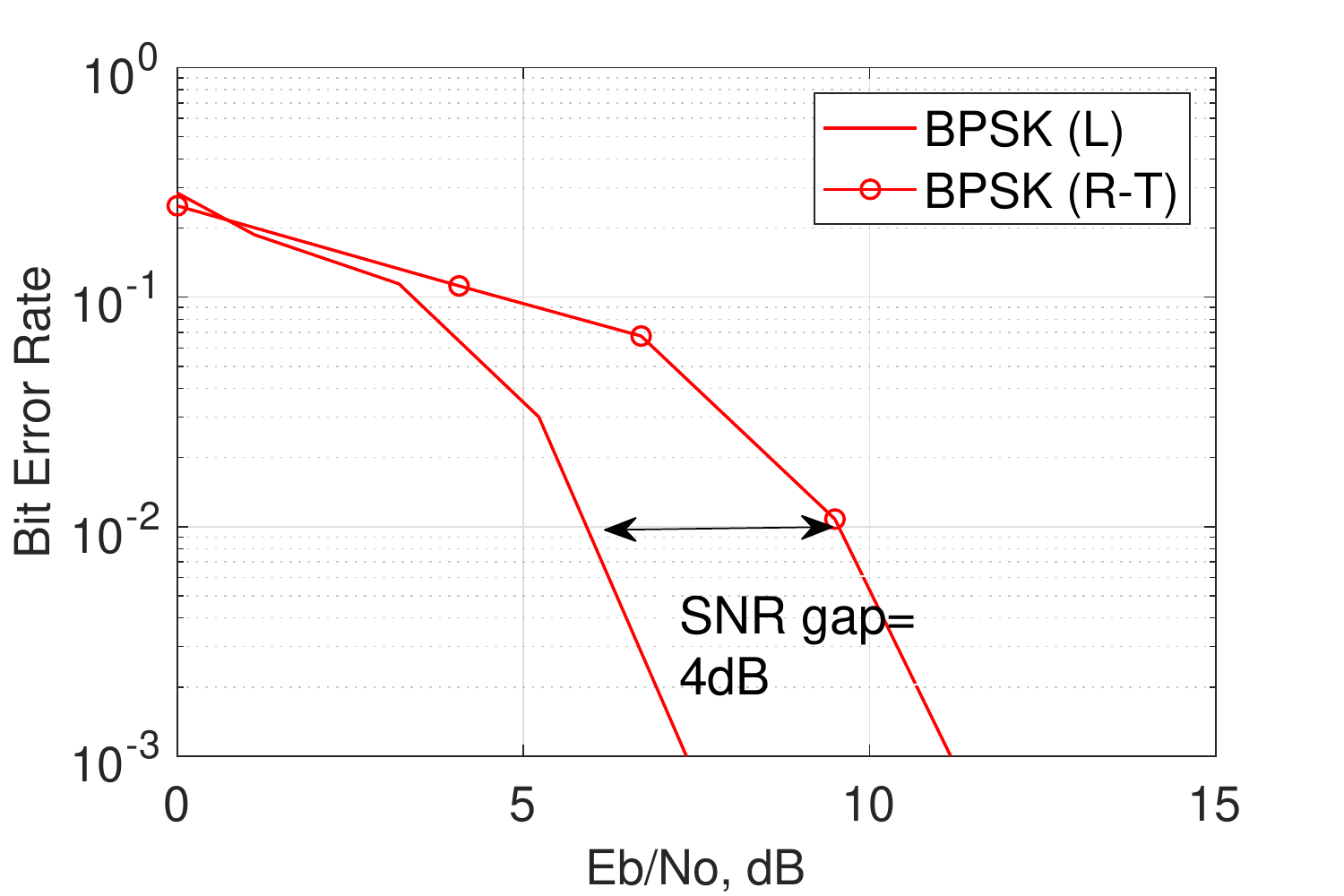}
        \caption{BPSK}
        \label{fig:gull}
    \end{subfigure}
    \quad
    \begin{subfigure}[b]{0.23\textwidth}
        \includegraphics[width=\textwidth]{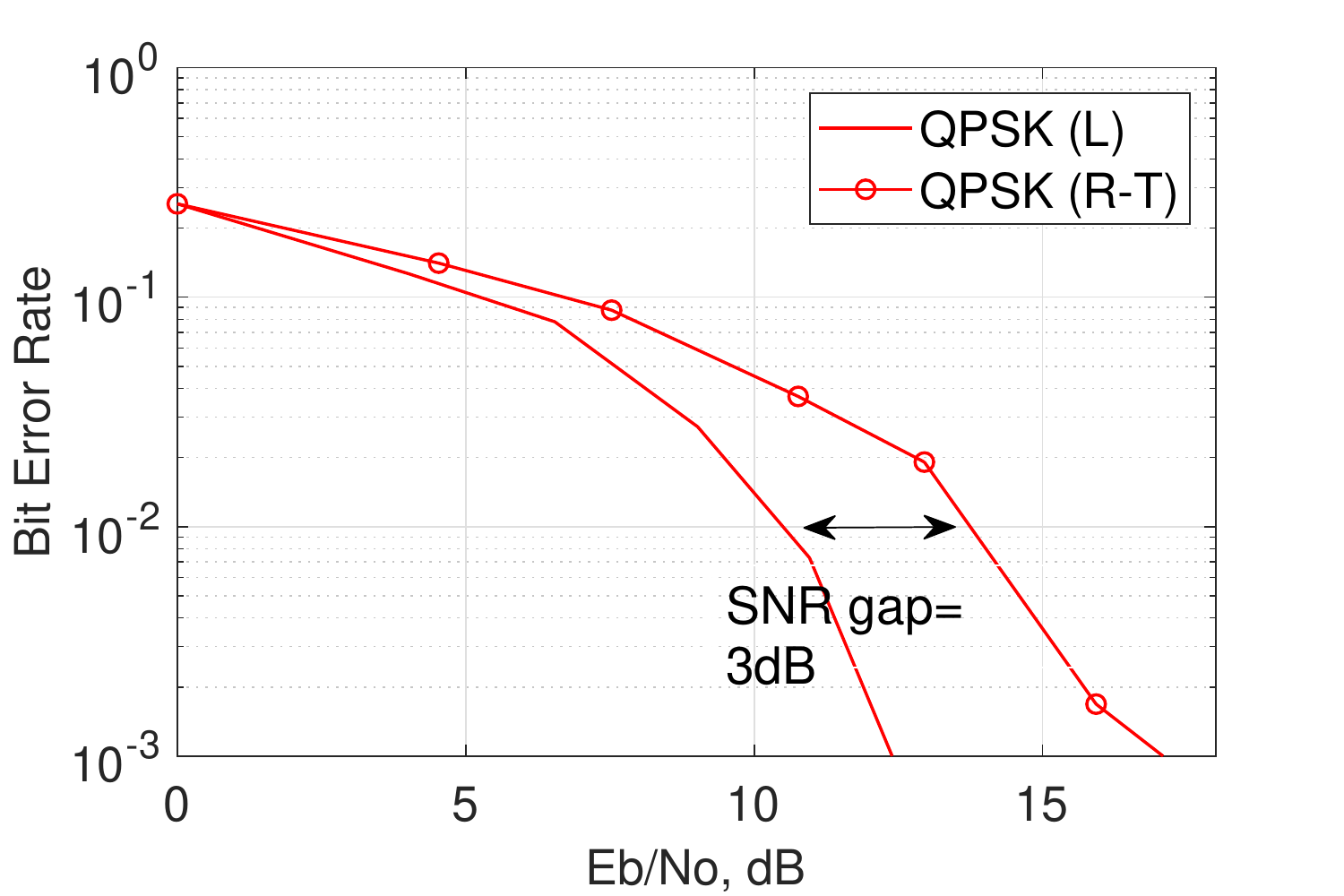}
        \caption{QPSK}
        \label{fig:tiger}
    \end{subfigure}
    \quad
    \begin{subfigure}[b]{0.23\textwidth}
        \includegraphics[width=\textwidth]{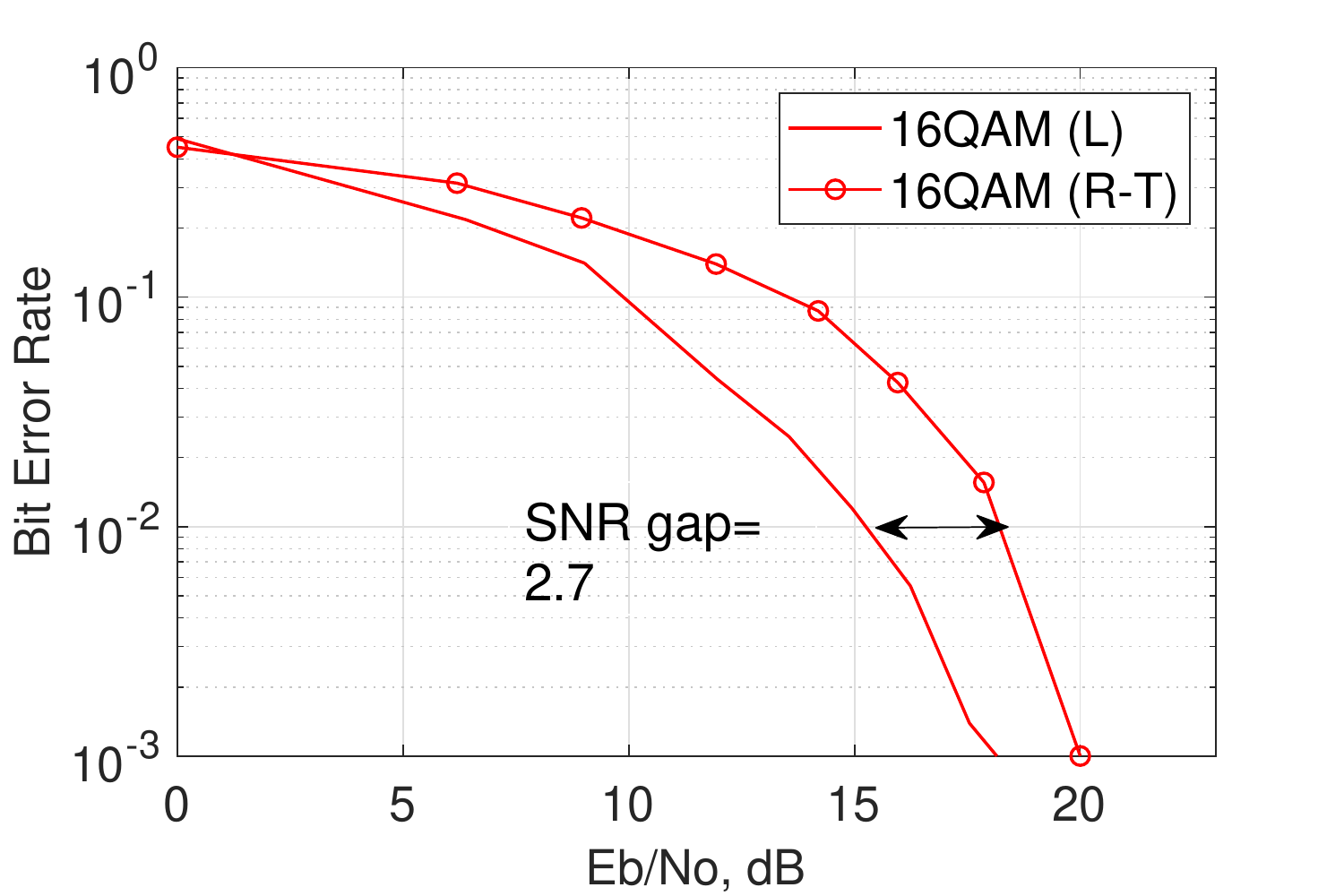}
        \caption{16QAM}
        \label{fig:gull}
    \end{subfigure}
    \quad
    \begin{subfigure}[b]{0.23\textwidth}
        \includegraphics[width=\textwidth]{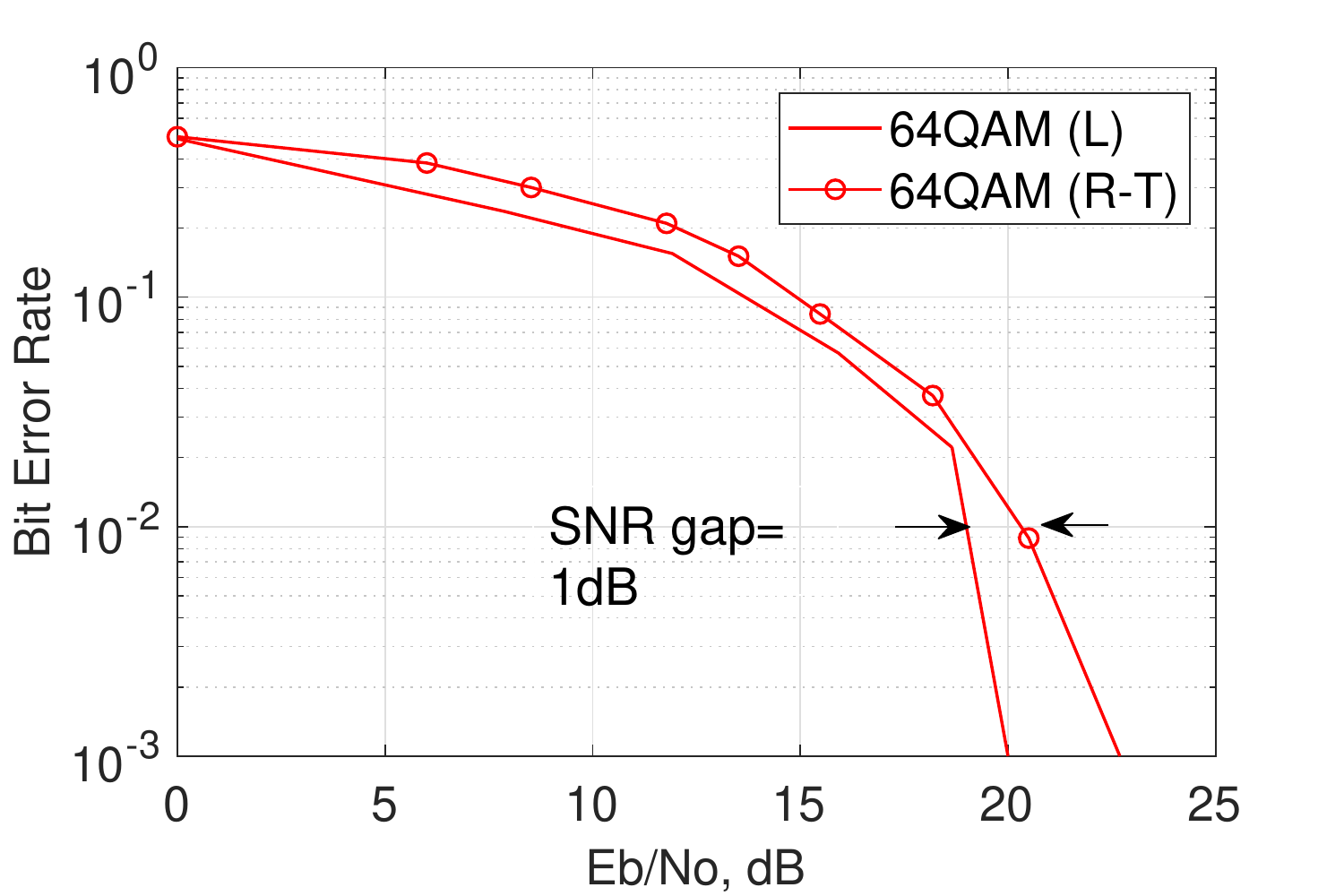}
        \caption{64QAM}
        \label{fig:gull}
    \end{subfigure}
    \caption{BER of {\abbrev} compared with legacy OFDM for over-the-air experiments using different modulation orders.}
    \label{BER_OTA}
  \vspace{-10pt}
\end{figure*}
\subsection{With Channel Estimation}
\label{subsec:perf_chEst}
\vspace{-1pt}
In this section, we analyze the performance of the channel estimation techniques, as elaborated in \S~\ref{subsec:channel} in two different channel models, `Flat Fading' and `Frequency Selective Fading', as shown in table~\ref{tbl:channel}.

Training signal is used to estimate the wireless channel effect for {\abbrev} transmissions, while legacy OFDM signal did not require this extra OFDM symbol. Figure~\ref{fig:train_flat} shows the performance of the training signal based channel estimation, which performs well due to the accuracy of the channel estimation using the training signal channel estimation. there is a small SNR penalty around $\approx 2dB$ due to the loss of orthogonality due to the randomization process at the transmitter.  This is expected as the channel effects are minor in a relative flat channel. Figure~\ref{fig:train_sel} shows the performance in a multipath rich environment, where the channel is extremely frequency selective. Results indicate that the newly introduced training signal is not only secured, but also provides accurate channel estimation for the signal to be reconstructed back with minimal SNR penalty. The SNR gap between legacy OFDM and {\abbrev} is $\approx 4dB$, indicating that it can be embraced in various practical scenarios.

\section{Over the Air experiments}
\label{sec:exp}
\begin{wrapfigure}{R}{0.2\textwidth}
\vspace{-15pt}
  \begin{center}
    \includegraphics[width=0.2\textwidth]{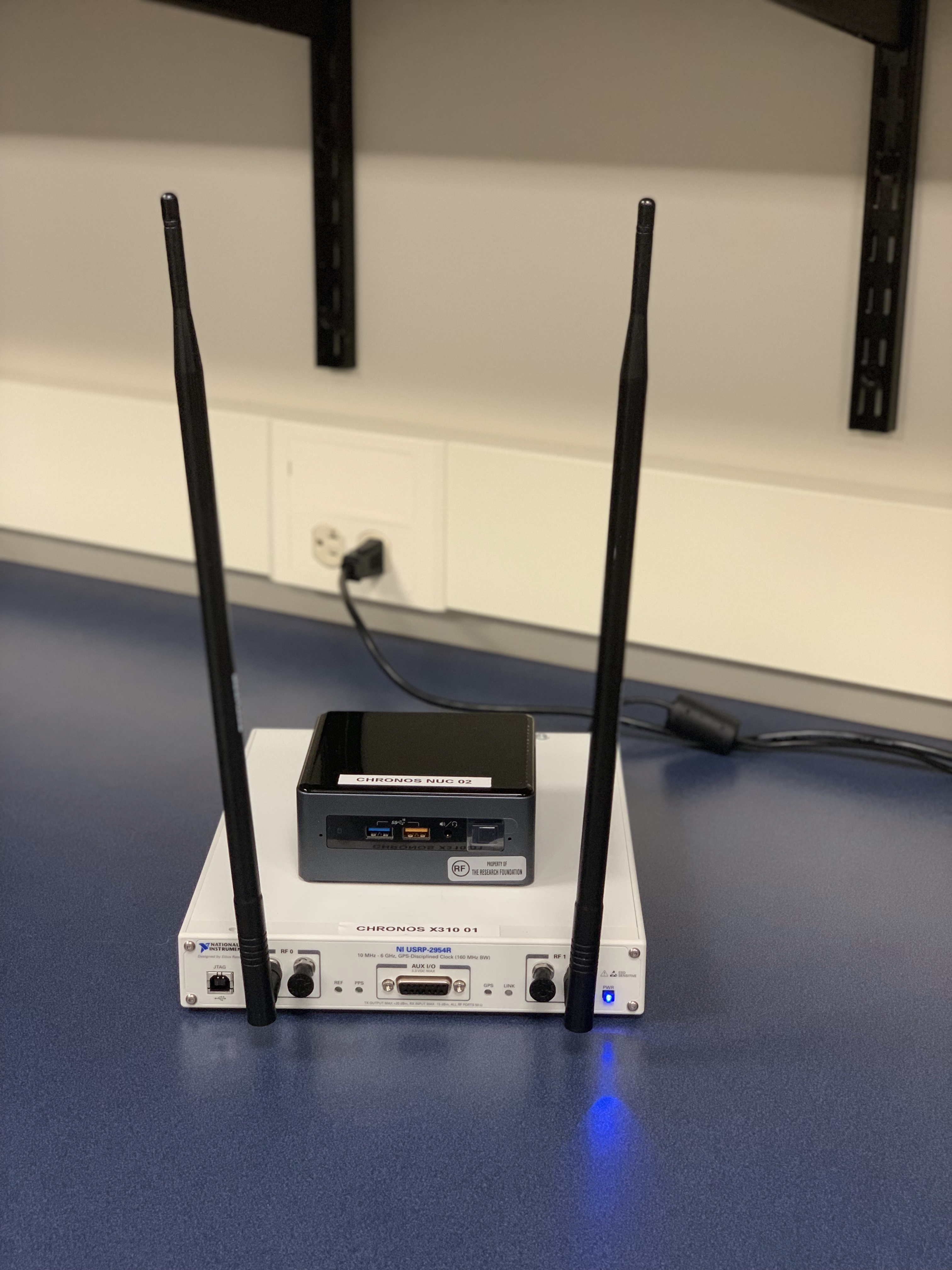}
  \end{center}
  \caption{Experimental setup of one node.}
\label{fig:experiment}
\vspace{-10pt}
\end{wrapfigure}

We perform extensive over-the-air experiments in indoor scenario to validate {\abbrev}. Figure~\ref{fig:experiment} shows one of our transceiver nodes equipped with USRP X310~\cite{X310} with 10 $dBi$ antenna. It is connected to an Intel NUC (NUC7i7BNH) with i7-7567U processor and 16GB DDR4 memory for faster processing of the I/O. The experiments are performed in multiple locations in a multipath-rich indoor environment in both line of sight and non-line-of-sight scenarios. We present the results for an average of all those locations to eliminate any dependencies on channel. Also, all the experiments were performed at 20MHz bandwidth and 2.484GHz frequency to avoid any interference from the Wi-Fi Access Points operating in the same area. Each data point in our result is an average of 500 OFDM symbols for both legacy OFDM and {\abbrev}.

Figure~\ref{BER_OTA} shows the BER performance for legacy and {\abbrev} transmissions for different modulation orders. It is evident that there is an SNR gap between the {\abbrev} and the Legacy OFDM transmission. This gap is due to the loss of orthogonal property of OFDM signal and channel estimation imperfections. This is the SNR penalty that we incur to secure a waveform in time domain.  Moreover, the SNR gap decreases at higher modulation order due to the higher operated SNR, which enables the receiver to decrease the error space in channel estimation using the training signal. In other words, the error introduced due to time-domain modification can be reconstructed back more efficiently at a higher SNR.  

\begin{figure}
    \centering
    \includegraphics[width=0.4\linewidth]{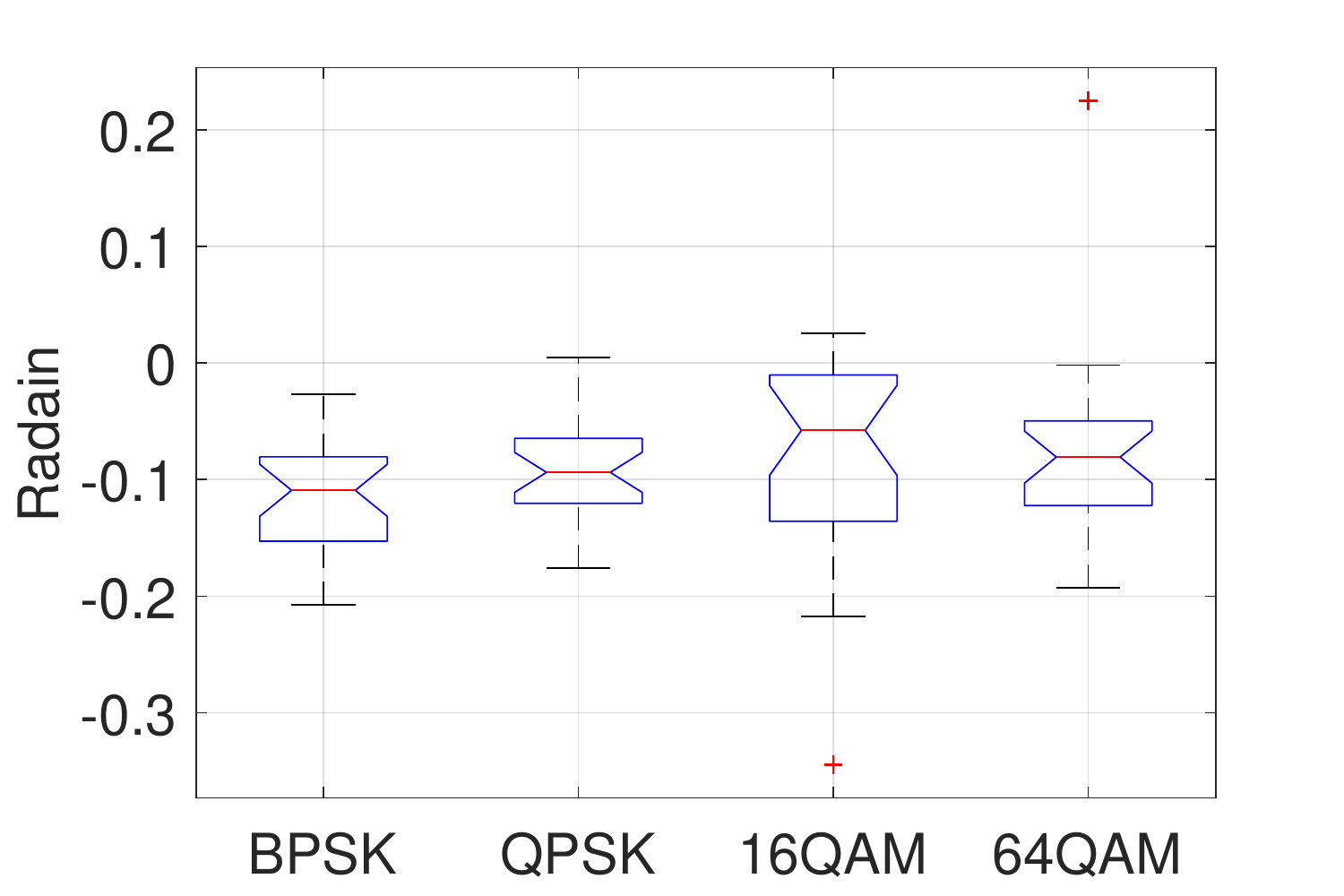}
    \caption{Residual phase offset correction for different modulation orders.}
    \label{fig:phase}
\end{figure}

Figure~\ref{fig:phase} presents the phase angle correction distribution for different modulation orders over all packets at different SNRs. The residual phase error is dependent on the hardware impairments and not on modulation order or channel characteristics. This is evident from the values, which varies between 0.05 to 0.12 radians. The phase correction results indicate that there exists significant residual error, which needs to be corrected to improve the performance.

\section{Conclusion}
\label{sec:conclude}

In this work, we present {\abbrev}, a secure OFDM transmission based on time domain scrambling using a shared secret key between the transmitter and the receiver. We perform channel estimation and equalization to retrieve the signal. Furthermore, we introduce a secured training signal to accurately estimate the channel followed by cluster based residual phase error correction. 
Over the air experiments show the success probability of this system. In future, the key generation and management can be developed based on channel state for lightweight secured physical layer encryption. 

\bibliographystyle{IEEEtran}
\bibliography{ref.bib}
\end{document}